\documentclass[12pt,twoside,dvips]{article}
\usepackage{amssymb}
\usepackage{amsfonts}
\usepackage{amsmath}
\usepackage{pifont}
\usepackage[spanish,english]{babel}
\usepackage{graphicx}
\usepackage{verbatim}

\begin{document}

\title{Production of multiple charged Higgs bosons in 3-3-1 models}
\author{R. Mart\'{\i}nez$\thanks{%
e-mail: remartinezm@unal.edu.co}$, F. Ochoa$\thanks{%
e-mail: faochoap@unal.edu.co}$, \and Departamento de F\'{\i}sica, Universidad
Nacional de Colombia, \\ Bogot\'{a} D.C, Colombia}
%EndAName
 
\maketitle

\begin{abstract}
We undertake the study of the charged Higgs bosons predicted by the model with gauge symmetry $SU(3)_c \otimes SU(3)_L \otimes U(1)_X$. By considering Yukawa mixing couplings between small ($\sim$ GeV) and large ($\sim$ TeV) scales, we show that the hypercharge-one $H_1^{\pm}$ and hypercharge-two $H_2^{\pm}$ Higgs bosons predicted by the model, can be simultaneously produced in $pp$ collisions at different production rates. At low energy, the $H_1^{\pm}$ bosons exhibit the same properties as the charged Higgs bosons from a two Higgs doublet model (2HDM), while $H_2^{\pm}$ are additional like-charged Higgs bosons from the underlying 3-3-1 model. Thus, the identification of multiple like-charged Higgs boson resonances may test the compatibility of theoretical models with experimental data. We study $H_{1,2}^{\pm}$ pair and associated $tbH_{1,2}^{\pm}$ productions at CERN LHC collider. In particular, we obtain that pair production can be as large as the single production in gluon-gluon collisions due to the interchange of a heavy neutral $Z'$ gauge boson predicted by the model. By considering decays to leptons $H_{1,2}^{\pm} \rightarrow \tau \nu _{\tau}$, we obtain scenarios where small peaks of $H_{2}^{\pm}$-boson events in transverse mass distributions can be identified over the $H_{1}^{\pm}$ background.                                
 
%We undertake a phenomenological study of the extra neutral Z' boson in the Minimal Quiver Standard Model and discuss limits on the model's parameters from previous precision electroweak experiments, as well as detection prospects at the Large Hadron Collider at CERN. We find that masses lower than around $700$ GeV are excluded by the $Z$-pole data from the CERN-LEP collider, and below $620$ GeV by experimental data from di-electron events at the Fermilab-Tevatron collider. We also find that at a mass of 1 TeV the LHC cross section would show a small peak in the di-lepton and top pair channel. 
\end{abstract}

\section{Introduction}

The mechanism that breaks the electroweak symmetry, which provides masses to all elementary particles is a subject of high priority in particle physics. In the Standard Model (SM) \cite{SM}, the inclusion of one $SU(2)_L$ scalar doublet leads to the symmetry breaking and generates the masses of the fundamental particles at the scale of the vacuum expectation value $\upsilon = 246$ GeV. One of the consequences of this mechanism is the prediction of one massive neutral scalar boson: the SM Higgs Boson (SMHB). Searches for Higgs bosons have been carried out by different experiments at CERN-LEP \cite{LEP}, Fermilab-Tevatron and CERN-LHC colliders. Recently, the ATLAS and CMS collaborations at LHC obtained evidence of an scalar resonance around the 126 GeV \cite{scalar_signal_ATLAS}, \cite{scalar_signal_CMS}. However, further analysis will be necessary to test the compatibility of this signal with the SMHB. Other searches have excluded the production of neutral Higgs bosons in other regions of the Higgs mass. At Fermilab-Tevatron, the region $156$ GeV $< m_{H} < 177$ GeV \cite{TEV} have been excluded. At CERN-LHC, the ATLAS collaboration have excluded the regions, $131-238$ GeV, and $251-466$ GeV \cite{ATLAS}, while the CMS collaboration have excluded the mass range $127-600$ GeV at 95 \% CL \cite{CMS}.  

Even if the existence of the SMHB is confirmed by the future experimental analysis, it seems "natural" that the SM must be extended to describe physics up to the Planck scale ($2.4\times 10^{18}$ GeV). On one hand, the mass of the SMHB is expected to be at the electroweak scale, as required from unitarity constraints \cite{unitarity}. However, on the other hand, if the limit of validity of the SM is near the Planck scale, there will arise huge quantum corrections to the squared Higgs mass, and "unnatural" cancellations must occur to give a Higgs boson of the order of the electroweak scale \cite{hierarchy}. Thus, it is to expect some sort of theory beyond the SM that fill the enormous range between the electroweak scale and the Planck scale. In particular, many theoretical extensions of the SM, charged Higgs bosons, $H^{\pm}$, are predicted. The detection of a charged Higgs boson could reveal many features about the underlying model beyond the SM. Searches at LEP in a general Two Higgs Doublet Model (THDM) \cite{type3} have set a limit $M_{H^{\pm}}>79.3$ GeV \cite{LEP-HPM}. For light charged Higgs bosons ($M_{H^{\pm}} \lesssim m_{top}$), in the framework of the Minimal Standard Supersymmetric Model (MSSM) \cite{MSSM}, the main production channel at the CERN-LHC is through top-quark decays $t\rightarrow bH^{+}$, while for heavy Higgs ($M_{H^{\pm}} \gtrsim m_{top}$), the associated $t\overline{b}$ production $pp\rightarrow tbH^{\pm}+X$ is the dominant mode \cite{LHC-report}. 

%On the other hand, in the 2HDM type III \cite{type3}, flavor changing neutral currents (FCNC) and CP-violating effects are predicted, which are severely suppressed by experimental data. One way to remove these effects, is by imposing discrete symmetries, obtaining 2HDM type I and type II \cite{2HDM}. Further, the general 2HDM can be classified according to the way in which the Yukawa mixing matrices rotate, generating 2HDM types $a$, $b$, $c$ and $d$ \cite{rodolfo}.
%Thus, if a 2HDM structure is confirmed by future experiments, the experimental data will restrict the parameters of the III-type, obtaining exact 2HDM type I, type II, or a low energy model type III restricted by approximate or slightly broken symmetries.
Other interesting alternative to extend the SM are the models with gauge symmetry $SU(3)_c \otimes SU(3)_L \otimes U(1)_X$, also called 3-3-1 models, which introduce a family nonuniversal $U(1)$ symmetry \cite{331-pisano, 331-frampton, 331-long, M-O}. These models have a number of phenomenological advantages. First of all, from the cancellation of chiral anomalies \cite{anomalias} and
asymptotic freedom in QCD, the 3-3-1 models can explain why there are three fermion
families. Secondly, since the third family is treated under a different representation, the
large mass difference between the heaviest quark family and the two lighter ones may be
understood \cite{third-family}. Finally, these models contain a natural Peccei-Quinn symmetry, necessary to solve the strong-CP problem \cite{PC}.

The 3-3-1 models extend the scalar sector of the SM into three $SU(3)_L$ scalar triplets: one heavy triplet field with vacuum expectation value (VEV) at high energy scale $\langle \chi \rangle = \nu _{\chi}$, which produces the breaking of the symmetry $SU(3)_L \otimes U(1)_X$ to the SM electroweak group $SU(2)_L \otimes U(1)_Y$, and two lighter triplets with VEVs at the electroweak scale  $\langle \rho \rangle = \upsilon _{\rho}$ and $\langle \eta \rangle = \upsilon _{\eta}$, which induce the electroweak breakdown. In the version without exotic charges, after the spontaneous breaking of the gauge symmetry and rotations into mass eigenstates, the model contains 4 massive charged Higgs ($H_1^{\pm}$, $H_2^{\pm}$), one neutral CP-odd Higgs ($A^0$), 3 neutral CP-even Higgs ($h^0, H^0_1, H_2^0$), and one complex neutral Higgs ($ H_3^0$) bosons. In particular, the charged sector is composed of two types of Higgs bosons: hypercharge-one Higgs bosons $H_1^{\pm}$ which exhibit tree-level couplings with the SM particles, and hypercharge-two Higgs $H_2^{\pm}$ bosons which show couplings with the SM matter through mixing with non-SM particles. An study of production of the hypercharge-one $H_1^{\pm}$ bosons at hadron colliders was performed by authors in ref. \cite{H1} in the framework of the 3-3-1 models.

In this work we show that the above 3-3-1 model can possess an specialized 2HDM type III at low energy, where an special basis is preferred, exhibiting specific couplings in the Yukawa constants \cite{quark-hierarchy}. Thus, like the 2HDM-III, the 3-3-1 model can predict huge flavor changing neutral currents (FCNC) and CP-violating effects, which are severely suppressed by experimental data at electroweak scales. One way to remove these effects, is by imposing discrete symmetries, obtaining two types of 3-3-1 models (type-I and -II models), which at low energy exhibit the same Yukawa interactions as the 2HDM type I and II. In the first case, one Higgs electroweak triplet (for example, $\rho$) provides masses to the phenomenological up- and down-type quarks, simultaneously. In the type-II model, one Higgs triplet ($\rho$) gives masses to the up-type quarks and the other triplet ($\eta$) to the down-type quarks. One way to distinguish a 2HDM from an underlying structure (3-3-1 triplet structures in our case) is by identifying multiple like-charged Higgs resonances, for example the two $H_{1,2}^{\pm}$ charged bosons. To explore this possibility, we show by the method of recursive expansion \cite{grimus} that if mixing couplings with the heavy quark sector of the 3-3-1 model is considered, then the hipercharge-two Higgs bosons $H_2^{\pm}$ can couple with the light SM quarks at tree-level. Thus, the $H_2^{\pm}$ Higgs bosons can be produced in the same production modes as the  $H_1^{\pm}$ Higgs, but at different statistical rates.  In particular, we study the $H_2^{\pm}$-boson production at CERN-LHC in $pp \rightarrow H_2^{+}H_2^{-}$ pair production and associated $tbH_2^{\pm}$ single production in the framework of the type-I and -II 3-3-1 models. For comparison purposes, we include the $H_1^{\pm}$-boson production for the above modes. We also discuss $H^{\pm}_{1,2}$ decay processes to tau leptons ($\tau$) assuming that $Br(H^{\pm}_{1,2}\rightarrow \tau \nu)=100$\%. We compare transverse mass distributions for different mass values of the Higgs bosons.

This paper is organized as follows. Section II is devoted to summarize the spectrum and interactions of the 2HDM and 3-3-1 models, in particular the interactions with charged Higgs bosons. We also show that, at low energy, the spectrum of the 3-3-1 model indeed can be separated into a light scale corresponding to the 2HDM type III, and a heavy scale decoupled from the light sector. In Section III we perfom the rotations into mass eigenstates of the quark sector taking into account small mixing terms between light and heavy scales. In Section IV, pair and associated production for charged Higgs bosons are studied in \textit{pp} collisions at LHC collider. Decay into tau leptons is considered in Sec. V. Finally in Sec. VI, we summarize our conclusions.

\section{The model}

In order to identify a 2HDM structure into the 3-3-1 model, we first show a brief summary of models with two Higgs doublets.
 
\subsection{2HDM-III Model}

The Yukawa's Lagrangian for the 2HDM type III is as follows

\begin{eqnarray}
-\mathcal{L}_Y&=&\overline{q^{i}_{L}}\left( \widetilde{\Phi}_1h_{1ij}^{U}+ \widetilde{\Phi}_2h_{2ij}^{U} \right){U}_{R}^{j}  \nonumber \\
&+&\overline{q^{i}_{L}}\left( \Phi_1 h_{1ij}^{D}+ \Phi_2 h_{2ij}^{D}\right){D}_{R}^{j}  \nonumber \\
                  &+& \overline{l^{i}_{L}} \left( \Phi_1 h_{1ij}^{e}+ \Phi_2 h_{2ij}^{e}\right){e}_{R}^{j}+h.c,
\label{yukawa}
\end{eqnarray}  
where $q^{i}_L=(U^{i},D^{i})_L$ are the up- and down-type quark doublets for each left-handed flavor component $U^{i}: (u,c,t)$ and $D^{i}:(d,s,b)$ for $i=1,2,3$, while $l^{i}_L=(\nu ^{i},e^{i})_L$ are the neutral and charged leptons for three flavors. The coupling constants $h_{nij}^{f}$ are the components of non-diagonal $3 \times 3$ matrices in the flavor space for $f=U,D,e$ and $n=1,2$. $\Phi_{1,2}$ are identical hypercharge-one Higgs doublets and $\widetilde{\Phi}_{1,2}=i\sigma_2 \Phi_{1,2}^*$ are conjugate fields with $\sigma_2$ the Pauli matrix. Assuming that both doublets acquire vacuum expectation values (VEV), $\langle \Phi_1\rangle_0=\upsilon_1$ and $\langle \Phi_2\rangle_0=\upsilon_2$, the doublets are written as follows:

\begin{eqnarray}
\Phi_1=
\begin{pmatrix}
\phi_1^{\pm}\\
\frac{1}{\sqrt{2}}(\upsilon_1 + \xi _1 \pm i \zeta_1 ) \\
\end{pmatrix},
\Phi_2=
\begin{pmatrix}
\phi_2^{\pm}\\
\frac{1}{\sqrt{2}}(\upsilon_2 + \xi _2 \pm i \zeta_2 ) \\
\end{pmatrix}.
\label{2hdm}
\end{eqnarray}
The mass eigenstates are related to weak eigenstates by:

\begin{eqnarray}
\begin{pmatrix}
G^{\pm}\\
H^{\pm}\\
\end{pmatrix}
=
R_{\beta}
\begin{pmatrix}
\phi_1^{\pm} \\
\phi_2^{\pm} \\
\end{pmatrix}, \hspace{0.3cm}
%\end{eqnarray}
%\begin{eqnarray}
\begin{pmatrix}
G^{o}\\
A^{o}\\
\end{pmatrix}
=
R_{\beta}
\begin{pmatrix}
\zeta _1 \\
\zeta_2 \\
\end{pmatrix}, \hspace{0.3cm}
%\end{eqnarray}
%\begin{eqnarray}
\begin{pmatrix}
H^{o}\\
h^{o}\\
\end{pmatrix}
=
R_{\alpha}
\begin{pmatrix}
\xi _1 \\
\xi _2 \\
\end{pmatrix},
\label{mass-scalar}
\end{eqnarray}
with:

\begin{eqnarray}
R_{\beta,\alpha}=\begin{pmatrix}
C_{\beta,\alpha} && S_{\beta,\alpha}\\
-S_{\beta,\alpha} && C_{\beta,\alpha}\\
\end{pmatrix},
\end{eqnarray}
where tan$\beta = \upsilon _2 / \upsilon _1$, and $\alpha$ is a mixing angle expressed as a function of parameters from the Higgs potential. In the most general 2HDM, the parameter tan$\beta$ is considered as unphysical, thus, it is necessary to redefine the Higgs doublets into a basis independent from $\beta$ \cite{beta-independent}. However, if some type of restriction (discrete symmetries, symmetries from bigger gauge models, etc,) is imposed, this parameter can be defined in an special basis, thus it will be a physical parameter. 

After the spontaneous symmetry breaking, the Yukawa Lagrangian in Eq. (\ref{yukawa}) gives mass matrices in the quark sector of the form

\begin{eqnarray}
M_{U}&=&\frac{1}{\sqrt{2}}\left(h^{U}_{1} \upsilon_1+ h^{U}_{2} \upsilon_2 \right),  \nonumber \\
M_{D}&=&\frac{1}{\sqrt{2}}\left(h^{D}_{1} \upsilon_1+ h^{D}_{2} \upsilon_2 \right).
\label{ND-mass}
\end{eqnarray}

The quark mass eigenstates can be obtained by unitary transformations of the left- and right-handed weak eigenstates: $U'_{L,R}=V_{L,R}^UU_{L,R}, D'_{L,R}=V_{L,R}^DD_{L,R}$, where the Cabibbo-Kobayashi-Maskawa (CKM) matrix is defined by $\kappa =V_L^UV_L^{D\dagger}$. Thus, the matrices in Eq. (\ref{ND-mass}) are diagonalized by the following bi-unitary transformations:

\begin{eqnarray}
m_{U}&=&\frac{1}{\sqrt{2}}V_L^{U\dagger}\left(h^{U}_{1} \upsilon_{1}+ h^{U}_{2} \upsilon_{2} \right)V_R^U,  \nonumber \\
m_{D}&=&\frac{1}{\sqrt{2}}V_L^{D\dagger}\left(h^{D}_{1} \upsilon_{1}+ h^{D}_{2} \upsilon_{2} \right)V_R^D,
\label{D-mass}
\end{eqnarray}
where the rotated Yukawa couplings $h'^U_{1,2}=V_L^{U\dagger}h^{U}_{1,2}V_R^U$ and $h'^D_{1,2}=V_L^{D\dagger}h^{D}_{1,2}V_R^D$ are not in general simultaneously diagonalized, leading to FCNC at tree level. However, it is possible to suppress these FCNC terms by demanding discrete symmetries in the Lagrangians, obtaining two different models:

\vspace{0.5cm}

\textit{\underline{Type I}}: Here, the couplings with one Higgs doublet (for example  $\Phi _2$) is removed from the Lagrangian, then only one Higgs doublet ($\Phi _1$) provides masses to the up- and down-type quarks, simultaneously.

\vspace{0.5cm}

\textit{\underline{Type II}}: in this case the Yukawa couplings are restricted so that one Higgs doublet (for example, $\Phi _1$) gives masses to the up-type quarks and the other doublet ($\Phi _2$) to the down-type quarks.

\vspace{0.5cm}

%Taking into account the above classification, we can from Eqs. (\ref{D-mass}) either solve for $(h^{U,0}_2,h^{D,0}_2)$ (for type I model) or for $(h^{U,0}_2,h^{D,0}_1)$ (for type II model):

%\begin{eqnarray}
%\text{type }(a): &h^{U}_{1}=\frac{\sqrt{2}}{\upsilon \text{C}_{\beta}}V_L^{U}m_UV_R^{U\dagger}-\text{T}_{\beta} h^{U}_2, \nonumber \\
%                & h^{D}_{1}=\frac{\sqrt{2}}{\upsilon \text{C}_{\beta}}V_L^{D}m_DV_R^{D\dagger}-\text{T}_{\beta} h^{D}_2 \nonumber \\
%\text{type }(b): &h^{U}_{2}=\frac{\sqrt{2}}{\upsilon \text{S}_{\beta}}V_L^{U}m_UV_R^{U\dagger}-\text{Cot}{\beta} h^{U}_1, \nonumber \\
%                & h^{D}_{1}=\frac{\sqrt{2}}{\upsilon \text{C}_{\beta}}V_L^{D}m_DV_R^{D\dagger}-\text{T}_{\beta} h^{D}_2,
%\label{rotation}
%\end{eqnarray}

Specifically, after the rotations to mass eigenstates, the model exhibits the following type-I and -II Lagrangians for the charged Higgs Bosons:

%Thus, in terms of $h_1^{U,D}$, the Lagrangian type $a$ reads:

\begin{eqnarray}
-\mathcal{L}^{2HDM-I}_{H^{\pm}}&=&\overline{U'_L}\left[ \frac{-\sqrt{2}m_D\kappa }{\upsilon }T_{\beta} \right]D'_RH^{+} +\overline{D'_L}\left[ \frac{-\sqrt{2}\kappa ^{\dagger}m_U}{\upsilon }T_{\beta} \right]U'_RH^{-}+h.c, \notag \\
-\mathcal{L}^{2HDM-II}_{H^{\pm}}&=&\overline{U'_L}\left[ \frac{\sqrt{2}m_D\kappa }{\upsilon }cot {\beta } \right]D'_RH^{+} +\overline{D'_L}\left[\frac{-\sqrt{2}\kappa ^{\dagger}m_U}{\upsilon }T_{\beta} \right]U'_RH^{-}+h.c,
\label{type-a-b}
\end{eqnarray}
where $\upsilon =\sqrt{\upsilon _1 ^2+\upsilon _2 ^2}=246$ GeV is the electroweak VEV.

%while the Lagrangian type $b$ in terms of $h_{1(2)}^{U(D)}$ reads:

%\begin{eqnarray}
 %-\mathcal{L}^b&=&\frac{1}{\upsilon \text{S}_{\beta}}\overline{U}m^{U}U\left(\text{S}_{\alpha} H^0+\text{C}_{\alpha}h^0\right)+\frac{1}{\upsilon \text{C}_{\beta}}\overline{D}m^{D}D\left(\text{C}_{\alpha} H^0-\text{S}_{\alpha}h^0\right) \notag \\
     %                     &&-\frac{1}{\sqrt{2}\text{S}_{\beta}}\left[\overline{U}{h^{U}_{1}}U-\text{T}_{\beta}\overline{D}{h^{D}_{2}}D\right]\left(\text{S}_{(\alpha-\beta)} H^0+\text{C}_{(\alpha-\beta)} h^0\right) \notag \\
      %              &&-\frac{i}{\upsilon \text{T}_{\beta}} \left[\overline{U}m^{U}\gamma _5U+\text{T}^2_{\beta}\overline{D}m^{D}\gamma _5D\right]A^0 \notag\\
       %                &&+\frac{i}{\sqrt{2}\text{S}_{\beta}}\left[\overline{U}h^{U}_1\gamma _5U+\text{T}_{\beta}\overline{D}h^{D}_2\gamma _5D\right]A^0 \notag\\
        %            &&-\frac{i}{\upsilon }\left[\overline{U}m^{U}\gamma _5U-\overline{D}m^{D}\gamma _5D\right]G^0 \notag\\
        %            &&-\frac{\sqrt{2}}{\upsilon \text{T}_{\beta}}\overline{U}\left[m^{U}KP_L+\text{T}^2_{\beta}Km^DP_R\right]DH^+\notag\\
        %            &&+\frac{1}{\text{S}_{\beta}}\overline{U}\left[h^{U}_1KP_L+\text{T}_{\beta}Kh^D_2P_R\right]DH^+\notag\\
           %         &&-\frac{\sqrt{2}}{\upsilon}\overline{U}\left[m^{U}KP_L-Km^DKP_R\right]DG^+ +h.c.
           %         \label{type-b}
%\end{eqnarray}

%The first Lagrangian (Eq. \ref{type-a}) can be separated in a 2HDM type I plus FCNC terms, while the second (Eq. \ref{type-b}) is a 2HDM type II plus FCNC interactions.

\subsection{3-3-1 Model}

\subsubsection{Physical spectrum}

We consider a 3-3-1 model where the electric charge is defined by:

\begin{eqnarray}
Q=T_3-\frac{1}{\sqrt{3}} T_8+X,
\end{eqnarray}
with $T_3=\frac{1}{2}Diag(1,-1,0)$ and $T_8=(\frac{1}{2\sqrt{3}})Diag(1,1,-2)$. In order to avoid chiral anomalies, the model introduces in the fermionic sector the following $(SU(3)_c, SU(3)_L,U(1)_X)$ left- and right-handed representations:

\begin{comment}
one $(3,3,1/3)$ quark triplet, two $(3,3^*,0)$ quark triplets and three $(1,3,-1/3)$ lepton triplets. For the right-handed sector, we introduce the following singlets in order to obtain Dirac charged fermions: three $(3^*,1,-1/3)$ down-type quarks, three $(3^*,1,2/3)$ up-type quarks, three $(1,1,-1)$ electron-type leptons. In addition we must introduce three $(3^*,1,Q_{J_{1,2},T_1})$ and three $(1,1,0)$ right-handed singlets associated to the new non-SM quarks and leptons, respectively.  In summary, we have the following representations free from chiral anomalies:
\end{comment}

\begin{eqnarray}
Q^{1}_L
&=&
\begin{pmatrix}
U^{1} \\
D^{1} \\
T^{1} \\
\end{pmatrix}_L:(3,3,1/3) ,
\left\{
\begin{array}{c}
U^{1}_R :(3^*,1,2/3) \\
D^{1}_R :(3^*,1,-1/3) \\
T^{1}_R :(3^*,1,2/3) \\
\end{array}
\right.  \nonumber \\
Q^{2,3}_L
&=&
\begin{pmatrix}
D^{2,3} \\
U^{2,3} \\
J^{2,3} \\
\end{pmatrix}_L:(3,3^*,0), \left\{
\begin{array}{c}
D^{2,3}_R :(3^*,1,-1/3) \\
U^{2,3}_R :(3^*,1,2/3) \\
J^{2,3}_R :(3^*,1,-1/3) \\
\end{array}
\right. \nonumber  \\
L^{1,2,3}_L
&=&
\begin{pmatrix}
\nu ^{1,2,3} \\
e^{1,2,3} \\
(\nu^{1,2,3})^c \\
\end{pmatrix}_L :(1,3,-1/3) ,\left\{
\begin{array}{c}
e^{1,2,3}_R :(1,1,-1) \\
N_R^{1,2,3}:(1,1,0) \\
\end{array}
\right.,
\label{fermion_spectrum}
\end{eqnarray}
where $U^{i}_L$ and $D^{i}_L$ for $i=1,2,3$ are three up- and down-type quark components in the flavor basis,  while $\nu ^{i}_L$ and $e^{i}_L$ are the neutral and charged lepton families. The right-handed sector transforms as singlets under $SU(3)_L$ with $U(1)_X$ quantum numbers equal to the electric charges. In addition, we see that the model introduces heavy fermions with the following properties: a single flavor quark  $T^{1}$ with electric charge $2/3$, two flavor quarks $J^{2,3}$ with charge $-1/3$, three neutral Majorana leptons $(\nu^{1,2,3})^c_L$ and three right-handed Majorana leptons $N^{1,2,3}_R$.  On the other hand, the scalar sector introduces one triplet field with VEV $\langle \chi \rangle_0=\upsilon_{\chi}$, which provides the masses to the new heavy fermions, and two triplets with VEVs $\langle \rho \rangle_0=\upsilon_{\rho}$ and $\langle \eta \rangle_0=\upsilon_{\eta}$, which give masses to the SM fermions at the electroweak scale. The $(SU(3)_L,U(1)_X)$ group structure of the scalar fields are:

\begin{eqnarray}
\chi&=&
\begin{pmatrix}
\chi_1^{0}\\
\chi_2^{-} \\
\frac{1}{\sqrt{2}}(\upsilon_{\chi} + \xi _{\chi} \pm i \zeta_{\chi} ) \\
\end{pmatrix}: (3,-1/3) \notag \\
\rho&=&
\begin{pmatrix}
\rho_1^{+} \\
\frac{1}{\sqrt{2}}(\upsilon_{\rho} + \xi _{\rho} \pm i \zeta_{\rho} ) \\
\rho _3^{+} \\
\end{pmatrix} : (3,2/3) \notag \\
\eta &=&
\begin{pmatrix}
\frac{1}{\sqrt{2}}(\upsilon_{\eta} + \xi _{\eta} \mp i \zeta_{\eta} ) \\
\eta _2^{-} \\
\eta _3^{0}
\end{pmatrix}:(3,-1/3).
\label{331-scalar}
\end{eqnarray}

After the symmetry breaking, it is found that the mass eigenstates are related to the weak states in the scalar sector by \cite{331-long, M-O}:

\begin{eqnarray}
\begin{pmatrix}
G_1^{\pm}\\
H_1^{\pm}\\
\end{pmatrix}
=R_{\beta_T}\begin{pmatrix}
\rho ^\pm _1 \\
\eta ^\pm _2 \\
\end{pmatrix}&,& \hspace{0.3cm}
\begin{pmatrix}
G^{0}_1\\
A^{0}_1\\
\end{pmatrix}
=R_{\beta_T}\begin{pmatrix}
\zeta _{\rho} \\
\zeta _{\eta}\\
\end{pmatrix}, \hspace{0.3cm}
\begin{pmatrix}
H^{0}_1\\
h^{0}\\
\end{pmatrix}
=R_{\alpha_T}\begin{pmatrix}
\xi _{\rho} \\
\xi _{\eta} \\
\end{pmatrix} \label{331-mass-scalar-a}  \\
\begin{pmatrix}
G_2^{0}\\
H_2^{0}\\
\end{pmatrix}
=R\begin{pmatrix}
\chi ^0 _1 \\
\eta ^0 _3 \\
\end{pmatrix}&,& \hspace{0.3cm}
\begin{pmatrix}
G^{0}_3\\
H^{0}_3\\
\end{pmatrix}
=\frac{-1}{\sqrt{2}}R\begin{pmatrix}
\zeta _{\chi} \\
\xi _{\chi}\\
\end{pmatrix}, \hspace{0.3cm}
\begin{pmatrix}
G^{\pm}_2\\
H^{\pm}_2\\
\end{pmatrix}
=R\begin{pmatrix}
\chi _{2}^\pm \\
\rho _{3}^\pm \\
\end{pmatrix},
\label{331-mass-scalar-b}
\end{eqnarray}
with:

\begin{eqnarray}
R_{\beta_T,\alpha_T}=\begin{pmatrix}
C_{\beta_T} && S_{\beta_T}\\
-S_{\beta_T} && C_{\beta_T}\\
\end{pmatrix}, \hspace{0.3cm}
R=\begin{pmatrix}
-1 && 0\\
0 && 1\\
\end{pmatrix},
\end{eqnarray}
%\begin{eqnarray}
%\begin{pmatrix}
%G_1^{\pm}\\
%H_1^{\pm}\\
%\end{pmatrix}
%&=&
%\begin{pmatrix}
%C_{\beta_T} && S_{\beta_T}\\
%-S_{\beta_T} && C_{\beta_T}\\
%\end{pmatrix}
%\begin{pmatrix}
%\rho ^\pm _1 \\
%\eta ^\pm _2 \\
%\end{pmatrix}, \nonumber \\
%\begin{pmatrix}
%G_2^{\pm}\\
%H_2^{\pm}\\
%\end{pmatrix}
%&=&
%\begin{pmatrix}
%-1 && 0\\
%0 && 1\\
%\end{pmatrix}
%\begin{pmatrix}
%\chi ^\pm _1 \\
%\eta ^\pm _3 \\
%\end{pmatrix} \nonumber \\
%\begin{pmatrix}
%G^{0}_1\\
%A^{0}_1\\
%\end{pmatrix}
%&=&
%\begin{pmatrix}
%C_{\beta _T} && S_{\beta _T}\\
%-S_{\beta _T} && C_{\beta _T}\\
%\end{pmatrix}
%\begin{pmatrix}
%\zeta _{\rho} \\
%\zeta _{\eta}\\
%\end{pmatrix}, \nonumber \\
%\begin{pmatrix}
%G^{0}_2\\
%H^{0}_2\\
%\end{pmatrix}
%&=&\frac{-1}{\sqrt{2}}
%\begin{pmatrix}
%-1 && 0 \\
%0 && 1 \\
%\end{pmatrix}
%\begin{pmatrix}
%\zeta _{\chi} \\
%\epsilon _{\chi}\\
%\end{pmatrix}, \nonumber \\
%\begin{pmatrix}
%H^{0}_1\\
%h^{0}\\
%\end{pmatrix}
%&=&
%\begin{pmatrix}
%C_{\alpha _T} && S_{\alpha _T}\\
%-S_{\alpha _T} && C_{\alpha _T}\\
%\end{pmatrix}
%\begin{pmatrix}
%\epsilon _{\rho} \\
%\epsilon _{\eta} \\
%\end{pmatrix} \nonumber \\
%\begin{pmatrix}
%G^{0}_3\\
%H^{0}_3\\
%\end{pmatrix}
%&=&
%\begin{pmatrix}
%-1 && 0 \\
%0 && 1 \\
%\end{pmatrix}
%\begin{pmatrix}
%\chi _{2}^0 \\
%\rho _{3}^0 \\
%\end{pmatrix}
%\label{331-mass-scalar}
%\end{eqnarray}
where tan$\beta _T=\upsilon _\eta / \upsilon _\rho$, and $\alpha _T$ is a mixing angle obtained from the Higgs potential.\\

For the boson-vector spectrum, we are just interested in the physical neutral sector that corresponds to the photon $A$, the neutral weak boson $Z$ and a new neutral boson $Z'$,  which are written in terms of the electroweak $SU(3)_L \otimes U(1)_X$ gauge fields as \cite{331-long}, \cite{M-O}:

\begin{eqnarray}
 A_\mu&=&S_{W} W^3_{\mu}+ C_{W}\left(\frac{1}{\sqrt{3}}T_{W} W_{\mu}^8 + \sqrt{1-\frac{1}{3}(T_{W})^2}B_\mu \right), \notag \\
 Z_\mu&=&C_{W} W^3_{\mu} - S_{W}\left(\frac{1}{\sqrt{3}}T_{W} W_\mu^8 + \sqrt{1-\frac{1}{3}(T_W)^2}B_\mu \right), \notag \\
 Z'_\mu&=&-\sqrt{1-\frac{1}{3}(T_{W})^2}W_\mu^8 + \frac{1}{\sqrt{3}}T_{W} B_\mu, \notag
\label{vector-spectrum}
\end{eqnarray}

\noindent where the Weinberg angle is defined as %\cite{M-O}

\begin{equation}
S_{W}=Sin(\theta _W)=\frac{\sqrt{3}g_{X}}{\sqrt{3g_{L}^{2}+4g_{X}^{2}}},
\label{weinberg_angle}
\end{equation}
with $g_{L}$ and $g_{X}$ the coupling constants of the groups $SU(3)_{L}$ and $U(1)_{X}$, respectively.

\subsubsection{Couplings}

\noindent  Using the fermionic content from Eq. (\ref{fermion_spectrum}), the neutral gauge interactions for SM quarks reads \cite{M-O}:

\begin{equation}
\mathcal{L}_{D}^{NC}=eQ_{q}\overline{q}%
A\hspace{-0.2cm}/q+\frac{g_{L}}{2C_{W}}\overline{q}\left[ \gamma _{\mu
}\left( g_{v}^{q}-g_{a}^{q}\gamma _{5}\right) Z^{\mu }+\gamma
_{\mu }\left( \widetilde{g}_{v}^{q}-\widetilde{g}_{a}^{q}\gamma _{5}\right)
Z^ {\prime \mu}\right]q ,  \label{L-neutro}
\end{equation}

\noindent where $q$ is $U=(U^1,U^2,U^3)$ or $D=(D^1,D^2,D^3)$ for up- and down-type
quarks, respectively, and $Q_q$ the electric charge in units of the positron charge $e$. The vector and
axial-vector couplings of the $Z$ and $Z'$ bosons are written in table \ref{EW-couplings} for each SM quark \cite{sher}.

\begin{table}[tbp]
\begin{center}
\renewcommand{\arraystretch}{2}
\scalebox{0.9}[0.9]{\begin{tabular}{|c|c|c|c|c|}
\hline
$Fermion$ & $g_{v}^{q}$ & $g_{a}^{q}$ & $\widetilde{g}_{v}^{q}$
& $\widetilde{g}_{a}^{q}$ \\ \hline\hline
$D^{1}$ & $-\frac{1}{2}+\frac{2}{3}S_{W}^{2}$ & $-\frac{1}{2}$ & $\frac{-1}{6}\sqrt{3-4S_W^2}$
& $\frac{-1}{2\sqrt{3-4S_W^2}}$ \\ \hline
$D^{m}$ & $-\frac{1}{2}+\frac{2}{3}S_{W}^{2}$ & $-\frac{1}{2}$ & $\frac{3-2S_{W}^{2}}{6\sqrt{3-4S_{W}^{2}}}$ & $\frac{1-2S_{W}^{2}}{2\sqrt{3-4S_{W}^{2}}}$  \\ \hline
$U^{1}$ & $\frac{1}{2}-\frac{4}{3}S_{W}^{2}$ & $\frac{1}{2}$ & $\frac{-3-2S_{W}^{2}}{6\sqrt{3-4S_{W}^{2}}}$ & $\frac{-1+2S_{W}^{2}}{2\sqrt{3-4S_{W}^{2}}}$ \\ \hline
$U^{m}$ & $\frac{1}{2}-\frac{4}{3}S_{W}^{2}$ & $\frac{1}{2}$ & $\frac{3-8S_{W}^{2}}{6\sqrt{3-4S_{W}^{2}}}$ & $\frac{1}{2\sqrt{3-4S_{W}^{2}}}$  \\ \hline
\end{tabular}}
\end{center}
\caption{Vector and Axial couplings of SM quarks and Neutral Gauge Bosons. The index $m=2,3$ labels the $3^{\ast}$ multiplets.}
\label{EW-couplings}
\end{table}

On the other hand, from the kinetic term of the Higgs Lagrangian, we obtain the following Higgs-Higgs-Vector interaction associated with the charged Higgs sector:

\begin{eqnarray}
i\mathcal{L}^{HHV}&=&
-ie\left[
H^{+}_1H^{-}_1+H_{2}^{+}H_{2}^{-}\right] \left( p-q\right) ^{\mu }A_{\mu }
\notag \\
&-&\frac{ig_{L}}{2C_{W}}\left[C_{2W}H^{+}_1H^{-}_1
+2S_{W}^{2}H_{2}^{+}H_{2}^{-} \right] \left( p-q\right) ^{\mu }Z_{\mu }  \notag \\
&+&\frac{ig_{X}}{2\sqrt{3}T_{W}} \left[ \left( C_{2\beta _{T}}+T_{W}^{2}\right) H^{+}_1H^{-}_1
+2\left( 1+ T_{W}^{2}\right)
H_{2}^{+}H_{2}^{-}\right] \left( p-q\right) ^{\mu }Z_{\mu }^{\prime}.
\label{H-H-V-couplings}
\end{eqnarray}%

Finally, we obtain the following $SU(3)_L \otimes U(1)_X$ renormalizable Yukawa Lagrangian for the quark sector:

\begin{eqnarray}
-\mathcal{L}_Y &=& \overline{Q_L^{1}}\left(\eta h^{U}_{\eta 1j}+\chi h^{U}_{\chi 1j}\right)U_R^{j}+
\overline{Q_L^{1}}\rho h^{D}_{\rho 1j}D_R^{j}\notag \\
&+&\overline{Q_L^{1}} \rho h^{J}_{\rho 1m} J^{m}_R+\overline{Q_L^{1}}\left(\eta h^{T}_{\eta 11}+\chi h^{T}_{\chi 11}\right)T_R^{1} \notag \\
&+&\overline{Q_L^{n}}\rho ^* h^{U}_{\rho nj}U_R^{j}+\overline{Q_L^{n}}\left(\eta ^* h^{D}_{\eta nj}+\chi ^* h^{D}_{\chi nj}\right)D_R^{j} \notag \\
&+&\overline{Q_L^{n}}\left( \eta ^* h^{J}_{\eta nm}+\chi ^* h^{J}_{\chi nm}\right) J^{m}_R+\overline{Q_L^{n}} \rho^* h^{T}_{\rho n1}T_R^{1}
 + h.c,
 \label{331-yukawa}
\end{eqnarray}
where $n=2,3$ is the index that labels the second and third quark triplet shown in Eq. (\ref{fermion_spectrum}), and $h^{f}_{\phi ij}$ are the ${i,j}$ components of non-diagonal matrices in the flavor space associated with each scalar triplet $\phi : \eta , \rho, \chi$.

\subsubsection{Yukawa couplings at low energy}

We require the breakdown $SU(3)_L \times U(1)_X \rightarrow SU(2)_L \times U(1)_Y $ in the flavor sector. In order to identify the particle content at low energy, we use the branching rules shown in Tab. \ref{tab:BR}, where we identify the following $(SU(2)_L,U(1)_Y)$ left-handed (SM) doublet representations:

\begin{eqnarray}
q^{1,2,3}_L
&=&
\begin{pmatrix}
U^{1,2,3} \\
D^{1,2,3} \\
\end{pmatrix}_L:(2,1/3) ,
 \nonumber  \\
l^{1,2,3}_L
&=&
\begin{pmatrix}
\nu ^{1,2,3} \\
e^{1,2,3} \\
\end{pmatrix}_L :(2,-1) ,
\label{subfermion_spectrum}
\end{eqnarray}
while $T^{1}_L,J^{2,3}_L$ and $(\nu^{1,2,3})^c_L$ are $SU(2)_L$ singlets (which we will call non-SM fermions). The right-handed fermions are decomposed into $SU(2)_L$ singlets with weak hypercharge equal to the electric charge. The scalar sector contains the following hypercharge-one subdoublets:

\begin{eqnarray}
\Phi _1&=&
\begin{pmatrix}
\rho_1^{+} \\
\frac{1}{\sqrt{2}}(\upsilon_{\rho} + \xi_{\rho} + i \zeta_{\rho} ) \\
\end{pmatrix} : (2,1) \notag \\
\widetilde{\Phi}_2 &=&
\begin{pmatrix}
\frac{1}{\sqrt{2}}(\upsilon_{\eta} + \xi_{\eta} - i \zeta_{\eta} ) \\
\eta _2^{-}
\end{pmatrix}:(2,-1).\notag \\
\Phi_3&=&
\begin{pmatrix}
\chi_1^{+}\\
\chi_2^0 \\
\end{pmatrix}: (2,1) ,
\label{21-scalar}
\end{eqnarray}
while $\chi _0 =(1/\sqrt{2})( \xi_{\chi} \pm i \zeta_{\chi}) $ and $\eta _3^0$ are hipercharge-zero singlets, and $\rho _3^{\pm}$ (which according with (\ref{331-mass-scalar-b}) are identified with the charged Higgs $H_2^{\pm}$ bosons) are hipercharge-two singlets. In (\ref{21-scalar}) we define the conjugate field as $\widetilde{\Phi}_{2}=i\sigma_2 \Phi_{2}^*$. In the above decompositions, the $U(1)_Y$ weak hypercharge, the $U(1)_X$ charge and the electric charge were related by:

\begin{equation}
Q=T_3+Y/2=T_3-\frac{1}{\sqrt{3}}T_8+X.
\end{equation}

\begin{table}[tbp]
\begin{center}
\renewcommand{\arraystretch}{1.4}
%\begin{equation*}
\begin{tabular}{l}
$(SU(3)_L,U(1)_X) \rightarrow (SU(2)_L,U(1)_Y)$  \\
 \hline \hline
%\begin{array}{l}
$Q^1_L:(3,\frac{1}{3}) \longrightarrow q^1_L:(2,\frac{1}{3})+T^1_L:(1,\frac{4}{3})$
%\end{array}
\\ \hline
%\end{array} $
%\begin{array}{l}
%\begin{array}{l}
$Q^{2,3}_L:(3^*,0) \longrightarrow i\sigma _2q^{2,3}_L:(2,\frac{1}{3})+J^{2,3}_L:(1,\frac{-2}{3}) $
\\ \hline
%\end{array}
%\begin{array}{l}
$L^{1,2,3}_L:(3,\frac{-1}{3}) \longrightarrow l^{1,2,3}_L:(2,-1)+(\nu^{1,2,3})^c_L:(1,0)$
\\ \hline
%\end{array}
$\chi :(3,\frac{-1}{3}) \longrightarrow \widetilde{\Phi } _3:(2,-1)+\chi _0:(1,0) $
 \\ \hline
%\end{array} 
%\begin{array}{l}
$\eta :(3,\frac{-1}{3}) \longrightarrow \widetilde{\Phi} _2:(2,-1)+\eta ^{0}_3:(1,0)$ 
\\ \hline
%\end{array} $ \\ 
%\begin{array}{l}
$\rho :(3,\frac{2}{3}) \longrightarrow \Phi _1:(2,1)+\rho _3^+:(1,2)$
%\end{array} $
\end{tabular}
%\end{equation*}
\end{center}
\caption{Branching rules for the $(SU(3)_L,U(1)_X) \rightarrow  (SU(2)_L,U(1)_Y) $ symmetry breaking.}
\label{tab:BR}
\end{table}

Taking into account the above branching rules, the Yukawa Lagrangian in Eq. (\ref{331-yukawa}) can be separated as follows:

\begin{eqnarray}
-\mathcal{L}_Y&=&\overline{q^{i}_{L}}\left( \widetilde{\Phi}_1h_{\rho ij}^{U}+\widetilde{\Phi}_2h_{\eta ij}^{U}+\widetilde{\Phi}_3h_{\chi ij}^{U}\right){U}_{R}^{j} \nonumber \\
&+&\overline{q^{i}_{L}}\left( \Phi_1h_{\rho ij}^{D}+\Phi _2h_{\eta ij}^{D}+\Phi _3h_{\chi ij}^{D}\right){D}_{R}^{j} \nonumber \\
&-&\mathcal{L}_Y(T^1,J^n),
\label{21-yukawa}
\end{eqnarray}  
where $\mathcal{L}_Y(T^1,J^n)$ are mixing terms with the non-SM fermions. The Lagrangian in (\ref{21-yukawa}) exhibits the same form as the 2HDM-III in Eq. (\ref{yukawa}) for the quark sector, plus extra mixing terms associated with the heavy scalar doublet $\Phi _3$ and the other non-SM particles. Thus, for $\upsilon _{\chi} \gg \upsilon _{\rho, \eta} $, the 3-3-1 model exhibits an effective 2HDM-III couplings in the quark sector at low energy. Furthermore, due to the nonuniversal $U(1)_X$ values exhibited by the spectrum in (\ref{fermion_spectrum}) and (\ref{331-scalar}), not all couplings between quarks and scalars are allowed by the gauge symmetry, which leads us to zero-texture of the Yukawa constants in Eq. (\ref{21-yukawa}) \cite{quark-hierarchy}. If we consider a low energy scenario in which the particles at the heavy scale decouple from those at light scales, the quark mass eigenstates may be obtained separately for each scale by unitary transformations of the left- and right-handed weak eigenstates: $U'_{L,R}=V^{U\dagger}_{L,R}U_{L,R}$,  $D'_{L,R}=V^{D\dagger}_{L,R}D_{L,R}$, and $J'_{L,R}=V^{J\dagger}_{L,R}J_{L,R}$, while the single flavor $T^1$ quark decouple from other components, obtaining that $T'^1_{L,R}=T^1_{L,R}$. In this limit, after the symmetry breaking, and using the VEVs from Eq. (\ref{331-scalar}), we obtain the following mass terms for the SM quarks:

\begin{eqnarray}
M_{U}&=&\frac{1}{\sqrt{2}}\left(h^{U}_{\rho } \upsilon _{\rho } + h^{U}_{\eta } \upsilon _{\eta } \right),  \nonumber \\
M_{D}&=&\frac{1}{\sqrt{2}}\left(h^{D}_{\rho} \upsilon _{\rho } + h^{D}_{\eta } \upsilon _{\eta } \right),
\label{mass-matrix}
\end{eqnarray}
which exhibit the same form as the matrices in Eq. (\ref{ND-mass}) but with the change $\upsilon _{1,2} \rightarrow \upsilon _{\rho, \eta}$ and $h_{1,2} \rightarrow h_{\rho, \eta}$. Then, using bi-unitary transformation analogous to (\ref{D-mass}), we obtain the diagonal mass matrices

\begin{eqnarray}
m_{U}&=&V_L^{U\dagger}M_UV_R^U=\frac{1}{\sqrt{2}}V_L^{U\dagger}\left(h^{U}_{\rho} \upsilon_{\rho}+ h^{U}_{\eta} \upsilon_{\eta} \right)V_R^U,  \nonumber \\
m_{D}&=&V_L^{D\dagger}M_DV_R^D=\frac{1}{\sqrt{2}}V_L^{D\dagger}\left(h^{D}_{\rho} \upsilon_{\rho}+ h^{D}_{\eta} \upsilon_{\eta} \right)V_R^D. %\nonumber \\
%m_{J}&=&V_L^{J\dagger}M_JV_R^J=\frac{1}{\sqrt{2}}V_L^{J\dagger}h^{J}_{\chi} \upsilon_{\chi}V_R^J, \nonumber \\
%m_{T}&=&\frac{1}{\sqrt{2}}h^{T}_{\chi} \upsilon_{\chi},
\label{diag-mass}
\end{eqnarray}

By requiring appropriate discrete symmetries, we can restrict the Higgs couplings in the Yukawa Lagrangian. In particular, we may generate type-I and -II models analogous to the 2HDM-I or -II if we require the following generalized discrete symmetries on the scalar and right-handed fermion fields:

\begin{eqnarray}
\text{Type I:}\hspace{0.5cm} \rho &\rightarrow & \rho ,\hspace{0.5cm} \eta \rightarrow -\eta ,\hspace{0.5cm} \chi \rightarrow -\chi 
\notag \\
U_R &\rightarrow &U_R, \hspace{0.5cm} D_R \rightarrow D_R \notag \\
T_R &\rightarrow &-T_R, \hspace{0.5cm} J_R \rightarrow -J_R \notag \\
N_R &\rightarrow &-N_R, \hspace{0.5cm} e_R \rightarrow e_R \notag \\ 
\text{Type II:}\hspace{0.5cm}\rho &\rightarrow & \rho ,\hspace{0.5cm} \eta \rightarrow -\eta ,\hspace{0.5cm} \chi \rightarrow -\chi  
\notag \\
U_R &\rightarrow &U_R, \hspace{0.5cm} D_R \rightarrow -D_R \notag \\
T_R &\rightarrow &-T_R, \hspace{0.5cm} J_R \rightarrow -J_R \notag \\
N_R &\rightarrow &-N_R, \hspace{0.5cm} e_R \rightarrow e_R.
\label{discrete_sym}              
\end{eqnarray}
In type-I models, one Higgs triplet ($\rho$) provides masses to both the up- and down-type quarks, while in type-II models the triplets $\rho$ and $\eta$ give masses to the up- and down-type quarks, respectively. Thus, eqs. (\ref{diag-mass}) becomes:   

\begin{eqnarray}
\text{Type I:}\hspace{0.5cm} m_{U(D)}&=&\frac{1}{\sqrt{2}}V_L^{U(D)\dagger}\left(h^{U(D)}_{\rho} \upsilon_{\rho} \right)V_R^{U(D)},  \nonumber \\
\text{Type II:}\hspace{0.5cm} m_{U(D)}&=&\frac{1}{\sqrt{2}}V_L^{U(D)\dagger}\left(h^{U(D)}_{\rho (\eta)} \upsilon_{\rho (\eta)} \right)V_R^{U(D)}. %\nonumber \\
%m_{J}&=&V_L^{J\dagger}M_JV_R^J=\frac{1}{\sqrt{2}}V_L^{J\dagger}h^{J}_{\chi} \upsilon_{\chi}V_R^J, \nonumber \\
%m_{T}&=&\frac{1}{\sqrt{2}}h^{T}_{\chi} \upsilon_{\chi},
\label{mass-I-II}
\end{eqnarray}

If $\nu _{\rho,\eta} =\nu _{1,2} $, the mass eigenstates in (\ref{mass-scalar}) are the same as the mass eigenstates in Eq. (\ref{331-mass-scalar-a}). In particular, we can identify the hypercharge-one Higgs bosons $H_1^{\pm}$ as 2HDM-like charged bosons, while the other charged Higgs bosons (hipercharge-two) $H_2^{\pm}$ are new charged scalar particles beyond the 2HDM.
%Thus, we obtain both the type I and type II models analogous to (\ref{rotation}):

%\begin{eqnarray}
%\text{type I}: &h^{U}_{\rho}=\frac{\sqrt{2}}{\upsilon \text{C}_{\beta_T}}V_L^{U}m_UV_R^{U\dagger}-\text{T}_{\beta_T} h^{U}_{\eta}, \nonumber \\
%                & h^{D}_{\rho}=\frac{\sqrt{2}}{\upsilon \text{C}_{\beta_T}}V_L^{D}m_DV_R^{D\dagger}-\text{T}_{\beta_T} h^{D}_{\eta} \nonumber \\
%\text{type II}: &h^{U}_{\eta}=\frac{\sqrt{2}}{\upsilon \text{S}_{\beta_T}}V_L^{U}m_UV_R^{U\dagger}-\text{Cot}{\beta_T} h^{U}_{\rho}, \nonumber \\
%                & h^{D}_{\rho}=\frac{\sqrt{2}}{\upsilon \text{C}_{\beta_T}}V_L^{D}m_DV_R^{D\dagger}-\text{T}_{\beta_T} h^{D}_{\eta},
%\label{rotation-331}
%\end{eqnarray}

\section{Yukawa Lagrangian with mixing couplings}

If we consider the complete Lagrangian in (\ref{331-yukawa}) (including couplings with non-SM fermions), we find the following mass terms after the symmetry breaking \cite{quark-hierarchy}:

\begin{eqnarray}
-\langle \mathcal{L}_Y \rangle =
\overline {\psi _L}M_{\psi}\psi _{R}+h.c.=
\left(\overline{U_L},\overline{T_L}\right)M_{UT}
\begin{pmatrix}
U_R \\
T_R
\end{pmatrix}+\left(\overline{D_L},\overline{J_L}\right)M_{DJ}
\begin{pmatrix}
D_R \\
J_R
\end{pmatrix}
+h.c,
\label{mass-yukawa}
\end{eqnarray}
where $U_{L,R}=(U^{1},U^{2},U^{3})_{L,R}$ are the left- and right-handed up-type quark flavor vectors,  $D_{L,R}=(D^{1},D^{2},D^{3})_{L,R}$ the corresponding down-type quark vectors, $J_{L,R}=(J^{2},J^{3})_{L,R}$ are two-dimensional vectors associated with the heavy quarks with electric charge $-1/3$ in (\ref{fermion_spectrum}) and $T_{L,R}$ is the single component of the heavy quark with charge $2/3$. The mass matrices have the following structures in the basis $(U,T)$ and $(D,J)$, respectively:  

\begin{eqnarray}
M_{\psi }=\begin{pmatrix}
M_{light} && f_{light} \\
G_{heavy} && \Lambda_{heavy} \\
\end{pmatrix}: %\hspace{0.3cm}
\left\{ \begin{matrix} M_{UT}=\begin{pmatrix}
M_U && k \\
K && M_T \\
\end{pmatrix} \vspace{0.3cm}
\\
M_{DJ}=\begin{pmatrix}
M_D && s \\
S && M_J \\
\end{pmatrix}
\end{matrix}\right.,
\label{mixing-mass}
\end{eqnarray}
where $M_{U}$, $k$, $K$, and $M_T$ are $3 \times 3$, $3 \times 1$, $1 \times 3$, and $1 \times 1$ matrices, respectively, while $M_{D}$, $s$, $S$, and $M_J$ are $3 \times 3$, $3 \times 2$, $2 \times 3$, and $2 \times 2$ matrices, respectively. The components of the above mass matrices are:

\begin{eqnarray}
M_{light}&=&M_{U(D)}=\frac{1}{\sqrt{2}}\left(h_{\rho}^{U(D)}\upsilon _{\rho}+h_{\eta}^{U(D)} \upsilon _{\eta} \right), \notag \\
\Lambda _{heavy}&=&M_{T(J)}=\frac{1}{\sqrt{2}}h_{\chi }^{T(J)}\upsilon _{\chi},\notag \\
f_{light}&=&k(s)=\frac{1}{\sqrt{2}}\left(h_{\rho}^{T(J)}\upsilon _{\rho}+h_{\eta}^{T(J)} \upsilon _{\eta} \right), \notag \\
G_{heavy}&=&K(S)=\frac{1}{\sqrt{2}}h_{\chi }^{U(D)}\upsilon _{\chi},
\label{UJ-mass}
\end{eqnarray}
where $M_{light}$ and $f_{light}$ are of the order of $\upsilon _{\rho}, \upsilon _{\eta} \sim$ GeV, while $\Lambda _{heavy}$ and $G_{heavy}$ are of the order $\upsilon _{\chi} \gg$ GeV.  
\begin{comment}
for $M_{UT}$, and

\begin{eqnarray}
M_D&=&\frac{1}{\sqrt{2}}\left(h_{\rho}^{D}\upsilon _{\rho}+h_{\eta}^{D} \upsilon _{\eta} \right), \hspace{0.5cm}  M_J=\frac{1}{\sqrt{2}}h_{\chi }^{J}\upsilon _{\chi},\notag \\
s&=&\frac{1}{\sqrt{2}}\left(h_{\rho}^{J}\upsilon _{\rho}+h_{\eta}^{J} \upsilon _{\eta} \right), \hspace{0.5cm}  S=\frac{1}{\sqrt{2}}h_{\chi }^{D}\upsilon _{\chi},
\label{DT-mass}
\end{eqnarray}

for $M_{DJ}$.
\end{comment}
The diagonalization of the matrices in (\ref{mixing-mass}) can be obtained by unitary transformations of the left- and right-handed weak eigenstates:

\begin{eqnarray}
\psi '_{L,R}=\mathcal{O}_{L,R}^{\dagger}\psi _{L,R},
\label{mass-states}
\end{eqnarray}
where $\psi ' _{L,R}=(U',T')_{L,R}$ or $(D',J')_{L,R}$ are the mass eigenstates. In order to find the form of the matrices $\mathcal{O}_{L,R}$, we separate them into two rotations:

\begin{eqnarray}
\mathcal{O}_{L,R}&=&\mathcal{U}_{L,R}W_{L,R}=\begin{pmatrix}
V_{L,R} && 0 \\
0 && P_{L,R}\\
\end{pmatrix}\begin{pmatrix}
1 && B_{L,R} \\
-B_{L,R}^{\dagger} && 1\\
\end{pmatrix},
\label{double-rotation}
\end{eqnarray} 
where $V_{L,R} = V_{L,R}^{U,D}$ are the same bi-unitary transformations as in (\ref{diag-mass}) that diagonalize the $U$ and $D$ components in the low energy limit, and $P_{L,R}=V_{L,R}^{J}$ or $1$ that rotate the $J$ or $T$ components, respectively. Since the first rotation through $\mathcal{U}_{L,R}$  does not lead to a completely diagonal mass matrix due to the mixing terms $f_{light}$ and $G_{heavy}$ in (\ref{mixing-mass}), we choose bi-unitary rotations $B_{L,R}=B_{L,R}^{U,D}$  by requiring the vanishing of the off-diagonal components of the following matrix:

\begin{eqnarray}
\mathcal{O}_{L}^{\dagger}M_{\psi }\mathcal{O}_R=W_{L}^{\dagger}\mathcal{U}_{L}^{\dagger}M_{\psi} \mathcal{U}_RW_R=\widetilde{m}_{\psi }=\begin{pmatrix}
\widetilde{m} _{light} && 0\\
0 && \widetilde{m} _{heavy}\\
\end{pmatrix}.
\label{diagonal-total}
\end{eqnarray}

Considering the scenario of small mixing terms (near the decoupling limit) ($G_{heavy} \ll \Lambda_{heavy}$, $f_{light} \ll M_{light}$) and using the method of recursive expansion \cite{grimus}, the authors in ref. \cite{quark-hierarchy} shows the following solutions:

\begin{eqnarray}
B_{L}&\approx& \left(m_{l}\widetilde{G}^{\dagger}+\widetilde{f}m_{h}^{\dagger}\right)\left(m_{h}^2 \right)^{-1}, \notag \\
B_{R}^{\dagger }&\approx& \widetilde{G}\left(m_{h}\right)^{-1},
\label{solution-L}
\end{eqnarray}
where $m_{l}=V_L^{\dagger}M_{light}V_R$, $m_{h}=P_L^{\dagger}\Lambda _{heavy}P_R$, $\widetilde{f}=V_L^{\dagger}f_{light}P_R$ and $\widetilde{G}=P_L^{\dagger}G_{heavy}V_R$. Replacing (\ref{solution-L}) in (\ref{double-rotation}) and (\ref{mass-states}), we obtain the following mass eigenstates:

\begin{eqnarray}
U'_L&=&(V_L^{U\dagger})U_L-(B_L^U)T_L \approx (V_L^{U\dagger})U_L-\left[\frac{m_U}{m_T^2}(V_R^UK)^{\dagger}+\frac{\widetilde{k} }{m_T} \right]T_L \notag \\
T'_L&=&(V_L^UB_L^U)^{\dagger}U_L+T_L \approx \left[\frac{m_U}{m_T^2}\widetilde{K} (V_L^U)^{\dagger}+\frac{\widetilde{k} ^{\dagger}}{m_T}V_L^{U\dagger} \right]U_L+T_L \notag \\
U'_R&=&(V_R^{U{\dagger}})U_R-(B_R^U)T_R \approx (V_R^{U\dagger})U_R-\left[\frac{\widetilde{K} }{m_T} \right]T_R \notag \\
T'_R&=&(V_R^UB_R^U)^{\dagger}U_R+T_R  \approx \left[\frac{\widetilde{K}}{m_T}V_R^{U\dagger} \right]U_R+T_R
\label{eigenmass-up}
\end{eqnarray}

\begin{eqnarray}
D'_L &=& (V_L^{D\dagger})D_L-(B_L^DV^{J\dagger}_L)J_L \approx (V_L^{D\dagger})D_L-\left[\frac{m_D}{m_J^2}(V_R^DS)^{\dagger}+\frac{\widetilde{s} }{m_J}V_L^{J\dagger} \right]J_L \notag \\
J'_L&=&(V_L^DB_L^D)^{\dagger}D_L+(V_L^{J\dagger})J_L \approx \left[\frac{m_D}{m_J^2}\widetilde{S} (V_L^D)^{\dagger}+\frac{\widetilde{s} ^{\dagger}}{m_J}V_L^{D\dagger} \right]D_L+(V_L^{J\dagger})J_L \notag \\
D'_R&=&(V_R^{D^{\dagger}})D_R-(B_R^DV_R^{J{\dagger}})J_R \approx (V_R^{D\dagger})D_R-\left[\frac{\widetilde{S} }{m_J}V_R^{J\dagger} \right]J_R \notag \\
J'_R&=&(V_R^DB_R^D)^{\dagger}D_R+(V_R^{J\dagger})J_R \approx \left[\frac{\widetilde{S}}{m_J}V_R^{D\dagger} \right]D_R+(V_R^{J\dagger})J_R.
\label{eigenmass-down}
\end{eqnarray} 
 
With the rotations from (\ref{eigenmass-up}) and (\ref{eigenmass-down}) for the quarks, and (\ref{331-mass-scalar-a})-(\ref{331-mass-scalar-b}) for the Higgs, the Yukawa Lagrangian of the 3-3-1 model in (\ref{331-yukawa}) can be written in mass eigenstates. In particular, for the couplings between the charged Higgs bosons and the $(U',D')$ mass eigenstates, we obtain:

\begin{eqnarray}
-\mathcal{L}_Y^{H^{\pm}_{1,2}}&=&\overline{U'_L}(V_L^{U\dagger})\left[ \left(-h_{\rho}^{D}S_{\beta _T}+h_{\eta}^{D} C_{\beta _T} \right)V_R^{D}-\left(-h_{\rho}^{J}S_{\beta _T}+h_{\eta}^{J} C_{\beta _T} \right)V_R^JB_R^{D\dagger} \right]D'_RH_1^{+} \notag \\
&+&\overline{D'_L}(V_L^{D\dagger})\left[ \left(-h_{\rho}^{U}S_{\beta _T}+h_{\eta}^{U} C_{\beta _T} \right)V_R^{U}-\left(-h_{\rho}^{T}S_{\beta _T}+h_{\eta}^{T} C_{\beta _T} \right)B_R^{U\dagger} \right]U'_RH_1^{-} \notag \\
&+&\overline{D'_L}(B_L^DV_L^{J\dagger})\left[ -h_{\rho}^{U}V_R^U+h_{\rho}^{T} B_R^{U\dagger} \right]U'_RH_2^{-} \notag \\
&+&\overline{U'_L}(B_L^{U\dagger})\left[ -h_{\rho}^{D}V_R^D+h_{\rho}^{J} B_R^{U\dagger} \right]D'_RH_2^{+} +h.c.
\label{Hcharged-coupling}
\end{eqnarray}

Since the rotations $B^{U,D}_{L,R}$ are suppressed by the inverse of the masses of the heavy $T$ and $J$ quarks according to (\ref{solution-L}) (where $m_h=m_{T,J}$), the couplings associated with $h^{J,T}_{\rho, \eta}$ in (\ref{Hcharged-coupling}) are negligible with respect to the $h^{D,U}_{\rho, \eta}$ couplings. In addition, other terms can be removed by the discrete symmetries in (\ref{discrete_sym}). Taking into account the above facts and using Eqs. (\ref{mass-I-II}), the Lagrangian (\ref{Hcharged-coupling}) for type-I and -II models can be written as:

\begin{eqnarray}
-\mathcal{L}_Y^{I}&=&\overline{U'_L}\left[ \frac{-\sqrt{2}m_D\kappa }{\upsilon }T_{\beta _T} \right]D'_RH_1^{+} +\overline{D'_L}\left[ \frac{-\sqrt{2}\kappa ^{\dagger}m_U}{\upsilon }T_{\beta _T} \right]U'_RH_1^{-} \notag \\
&+&\overline{D'_L}\left[ \frac{-\sqrt{2}m_U}{\upsilon  C_{\beta _T}} (B_L^DV_L^{J\dagger}V_L^U) \right]U'_RH_2^{-} +\overline{U'_L}\left[ \frac{-\sqrt{2}m_D}{\upsilon C_{\beta _T}} (B_L^{U\dagger}V_L^D) \right]D'_RH_2^{+} \notag \\
&+&h.c.,
\label{Hcharged-coupling-a}
\end{eqnarray}

\begin{eqnarray}
-\mathcal{L}_Y^{II}&=&\overline{U'_L}\left[ \frac{\sqrt{2}m_D\kappa }{\upsilon }cot{\beta _T} \right]D'_RH_1^{+} +\overline{D'_L}\left[ \frac{-\sqrt{2}\kappa ^{\dagger}m_U}{\upsilon }T_{\beta _T } \right]U'_RH_1^{-} \notag \\
&+&\overline{D'_L}\left[ \frac{-\sqrt{2}m_U}{\upsilon  C_{\beta _T}} (B_L^DV_L^{J\dagger}V_L^U) \right]U'_RH_2^{-} %+\overline{U'_L}\left[ \frac{\sqrt{2}m_D}{\upsilon  C_{\beta _T}} (B_L^{U\dagger}V_L^D) \right]D'_RH_2^{+}
+h.c.,
\label{Hcharged-coupling-b}
\end{eqnarray}
where $\kappa$ is the CKM matrix and $\upsilon =\sqrt{\upsilon _{\rho}^2+\upsilon _{\eta}^2 }=246$ GeV is the electroweak VEV. We see that the couplings of the $H_1^{\pm}$ bosons in the above Lagrangians are analogous to the $H^{\pm}$-boson couplings of the 2HDM in (\ref{type-a-b}). Thus, we identify the $H_1^{\pm}$ bosons with a 2HDM-like charged Higgs bosons (furthermore, if at low energy we neglect the coupling with the $Z'$ boson in Eq. (\ref{H-H-V-couplings}), we obtain that $H^{\pm}_{1} \rightarrow H^{\pm}$).

\begin{comment}
we observe that the matrices have the following general structure:

\begin{eqnarray}
M=\begin{pmatrix}
M_{light} && f_{light}  \\
G_{heavy} && \Lambda _{heavy}  \\
\end{pmatrix},
\label{mixing-matrix}
\end{eqnarray}

where $M_{light} \sim f_{light} \sim \upsilon_{\rho} \sim 246$ GeV, while $G_{heavy}\sim \Lambda_{heavy}\sim \upsilon _{\chi} \gg 246$ GeV. The above mass matrix can be diagonalized by a bi-unitary transformation: $\widetilde{m}=\mathcal{O}_L^{\dagger}M\mathcal{O}_R$. This transformation can be separated into two rotations: first, we can rotate through the bi-unitary transformations $V_{L,R} = V_{L,R}^{U,D}$ and $P_{L,R}=V_{L,R}^{J}$ defined by Eq. (\ref{diag-mass}) in the low energy limit. Second, since the first rotation does not lead to a completely diagonal matrix due to the mixing terms $f$ and $G$, we must perform another rotation through unitary matrices $B_{L,R}$. Thus, we separate the original rotation into:
 
\begin{eqnarray}
\mathcal{O}_{L,R}&=&\mathcal{U}_{L,R}W_{L,R}=\begin{pmatrix}
V_{L,R} && 0 \\
0 && P_{L,R}\\
\end{pmatrix}\begin{pmatrix}
1 && B_{L,R} \\
-B_{L,R}^{\dagger} && 1\\
\end{pmatrix},
\label{double-rotation}
\end{eqnarray}

where we require that:

\begin{eqnarray}
\mathcal{O}_{L}^{\dagger}M\mathcal{O}_R=W_{L}^{\dagger}\mathcal{U}_{L}^{\dagger}M\mathcal{U}_RW_R=\widetilde{m}=\begin{pmatrix}
\widetilde{m} _{light} && 0\\
0 && \widetilde{m} _{heavy}\\
\end{pmatrix}.
\label{diagonal-total}
\end{eqnarray}

\end{comment}

\section{Production of charged Higgs bosons}

Since the couplings found in (\ref{Hcharged-coupling-a}) and (\ref{Hcharged-coupling-b}) are proportional to the quark masses, the largest contribution comes from the top quark ($m_t \approx 173$ GeV). Thus, considering the couplings to the third up type family, the dominant contribution of the Yukawa interactions in (\ref{Hcharged-coupling-a}) and (\ref{Hcharged-coupling-b}) are:

\begin{eqnarray}
-\mathcal{L}_Y^{I(II)}&\approx &\frac{-\sqrt{2}m_tT_{\beta _T}}{\upsilon }\overline{b_L}t_RH_1^{-}-\frac{\sqrt{2}m_tB_L^D}{\upsilon C_{\beta _T}}\overline{b_L}t_RH_2^{-}+h.c.,
\label{bottom-top-a}
\end{eqnarray}
%\begin{eqnarray}
%-\mathcal{L}_Y^{II}&\approx &\frac{-\sqrt{2}m_tcot\beta _T}{\upsilon }\overline{b_L}t_RH_1^{-}+\frac{-\sqrt{2}m_tB_L^D }{\upsilon S_{\beta _T}}\overline{b_L}t_RH_2^{-}+h.c.,
%\label{bottom-top-b}
%\end{eqnarray}
where we consider $\kappa _{tb}\sim 1$ for the $(3,3)$ component of the CKM matrix. We are interested in comparing the production ratios between $H_1^{\pm}$ and $H_2^{\pm}$, then in the above Lagrangian we also assume for simplicity that $V^{U,D,J}_{L,R} \sim 1$ for the U, D and J quarks. It is to note that type-I and type-II models exhibit different Yukawa couplings in the down sector, but the same couplings in the up sector. Since we neglected the contribution from the down-type masses, we see in Eq. (\ref{bottom-top-a}) that both models possess identical couplings.

Taking into account the couplings in (\ref{L-neutro}), (\ref{H-H-V-couplings}) and (\ref{bottom-top-a}) for type-I and -II models, we may study the production of $H_{1,2}^{\pm}$ Higgs bosons. Fig. \ref{fig-1} shows different partonic processes for Higgs production in $pp$ collisions, where $(a)$ corresponds to quark-antiquark annihilation for pair production through vector electroweak bosons, while $(b)$-$(f)$ corresponds to associated production $tbH_{1,2}^{\pm}$ through gluon-gluon collisons ($(b)$-$(d)$) and quark-antiquark annihilation ($(e)$ and $(f)$). 

\subsection{Pair production}

\begin{figure}[b]
\centering \includegraphics[scale=0.45]{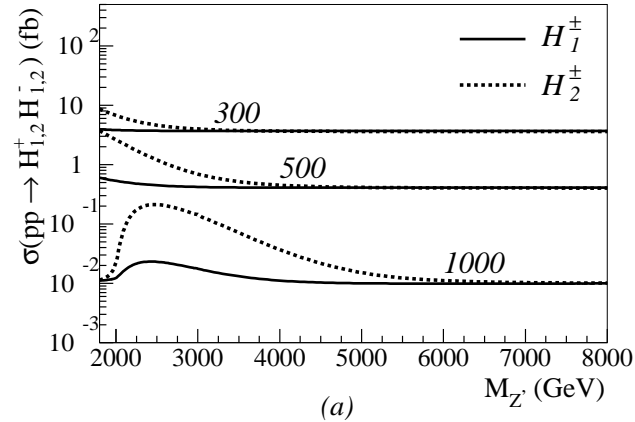}
\centering \includegraphics[scale=0.45]{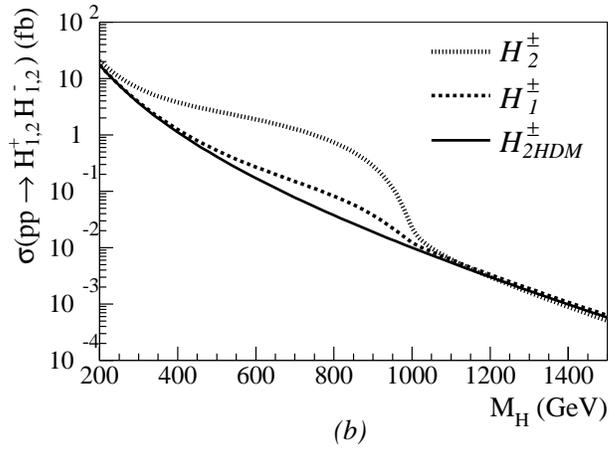}
\caption{Dependence of the cross section for charged Higgs-boson pair production with $(a)$ the $Z'$-boson mass for $M_{H}=300, 500$ and $1000$ GeV, and with $(b)$ the Higgs mass for $M_{Z'}=2$ TeV in 3-3-1 and 2HDM models.}
\label{fig-2}
\end{figure}

For the pair production from Fig. \ref{fig-1}$(a)$, we use the couplings in (\ref{H-H-V-couplings}). We observe that $T_{\beta _{T}}$, $M_{H_{1,2}}$ and $M_{Z'}$ are free parameters, while $g_L$ and $g_X$ can be parametrized as functions of the electric charge of the proton $e$ and the Weinberg angle: $e=g_LS_W$ and $g_X=\sqrt{3}g_LS_W/\sqrt{3-4S_W^2}$ (from definition in (\ref{weinberg_angle})). The parameter $T_{\beta _T}$ only appears in the coupling of $H_1^{\pm}$ with $Z'$ through the coefficient $(C_{2\beta _{T}}+T_W^2)$ in (\ref{H-H-V-couplings}). Just for numerical purposes we take $T_{\beta _{T}}=10$. At CM energy of 14 TeV in $pp$ collisions, we use the Calchep package \cite{calchep} to obtain the pair production cross sections for $H_1^{\pm}$ and $H_2^{\pm}$ bosons. Fig. \ref{fig-2}$(a)$ shows the dependence of the cross section with the mass of the $Z'$ boson for three different Higgs masses: $M_{H_{1,2}}=300, 500$ and $1000$ GeV. The behaviour shown by the curves can be understood as follows: For $M_{Z'} < 2M_{H_{1,2}}$, the production of real $Z'$ bosons is supressed by kinematical conditions, so that the contributions come only from $A$ and $Z$ bosons. Above the resonance region $M_{Z'}=2M_{H_{1,2}}$, the production of $Z'$ increases, which leads to larger cross sections. However, when the $Z'$ boson becomes heavier, the energy of the collision is not large enough to obtain an appreciable $Z'$ production, thus the $A$ and $Z$ contributions are again the dominant modes for pair production. On the other hand, we see that around the region of the $Z'$-resonance, the cross sections splits in two branches, where the $H_{2}^{\pm}$ bosons exhibit larger contributions than the 2HDM-like bosons $H_1^{\pm}$. This splits is due to the different contributions exhibits by the $H_1^{\pm}$ and $H_2^{\pm}$ bosons with the gauge bosons in Eqn. (\ref{H-H-V-couplings}). Using the notation $g_{H_iH_iV}$ to designate the Higgs-Higgs-Vector couplings, and with SM numerical inputs, we obtain the following relations:

\begin{eqnarray}
\frac{g_{H_2H_2A}}{g_{H_1H_1A}}&=&1 \notag \\
\frac{g_{H_2H_2Z}}{g_{H_1H_1Z}}&=&\frac{2S_{W}^{2}}{C_{2W}}\approx 0.8 \notag \\
\frac{g_{H_2H_2Z'}}{g_{H_1H_1Z'}}&=&\frac{2\left( 1+ T_{W}^{2}\right)}{\left( C_{2\beta _{T}}+T_{W}^{2}\right)}\approx 3.7,
\label{coupling_ratios}
\end{eqnarray}
which indicates that due to the $Z'$-couplings, the ratios between cross sections $\sigma({H_2^{\pm}})/\sigma({H_1^{\pm}})$ can be as large as $3.7^2 \sim 13.7$. i.e as one order of magnitude larger. Thus, we can see in Fig. \ref{fig-2}$(a)$ for the curves $M_{H_{1,2}}=1000$ GeV, that the $H_2^{\pm}$ boson exhibits a peak which is about one order of magnitude larger than the $H_1^{\pm}$ peak.   

On the other hand, Fig. \ref{fig-2}$(b)$ compares the dependence of the cross sections with the Higgs masses for $H_1^{\pm}$, $H_2^{\pm}$, and $H^{\pm}_{2HDM}$ Higgs bosons. We use the value $M_{Z'}=2$ TeV. As with the Fig. \ref{fig-2}$(a)$, the cross sections splits into various branches for $M_{H_{1,2}} \le M_{Z'}/2 = 1$ TeV, where the largest contribution comes from the $H_2^{\pm}$ bosons. In particular, we see that for 2HDM we obtain smaller production ratios due to the fact that in this model the $Z'$ contribution does not exist.

\subsection{Single production}

\begin{figure}[b]
\centering \includegraphics[scale=0.45]{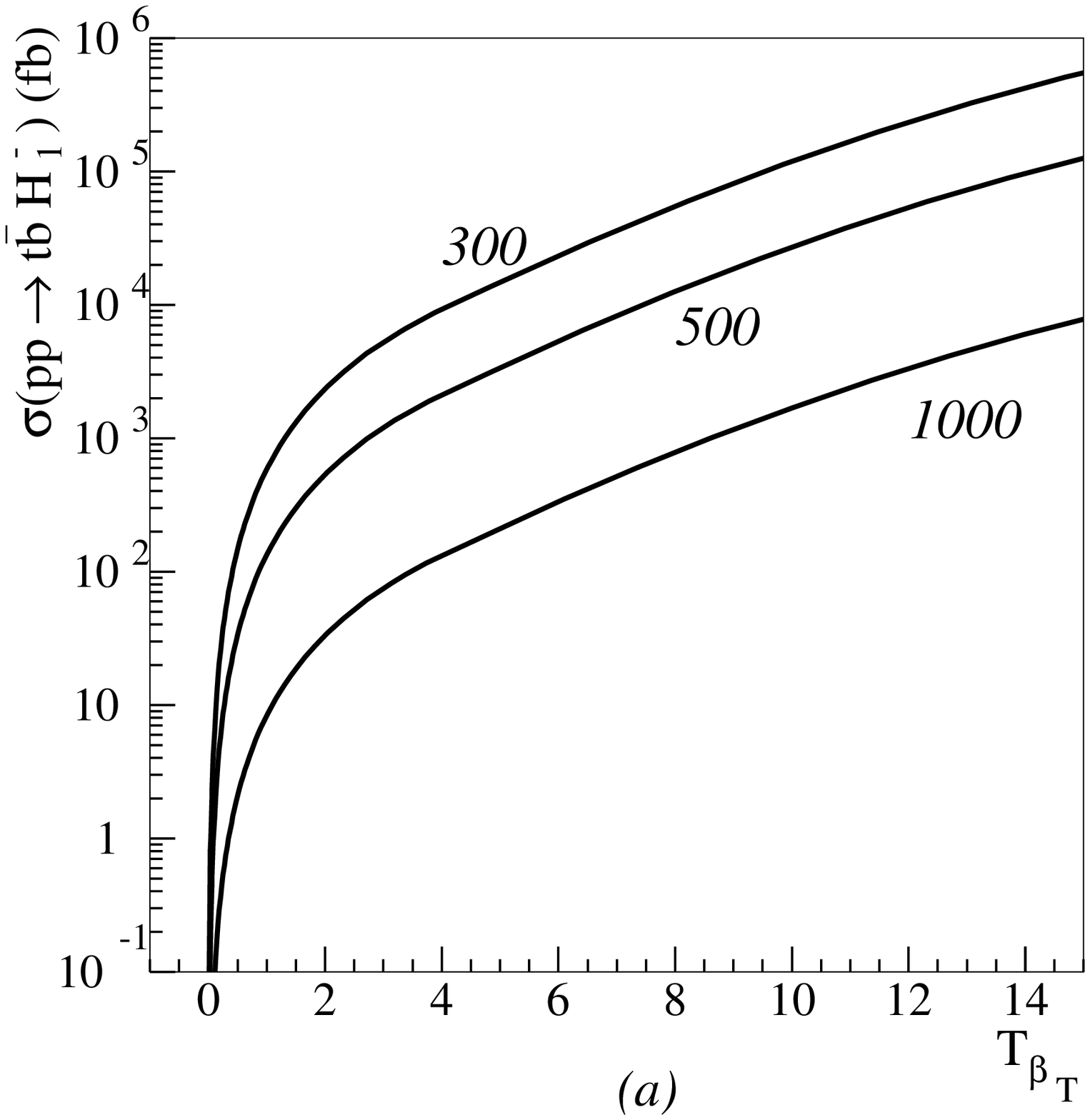}
\centering \includegraphics[scale=0.45]{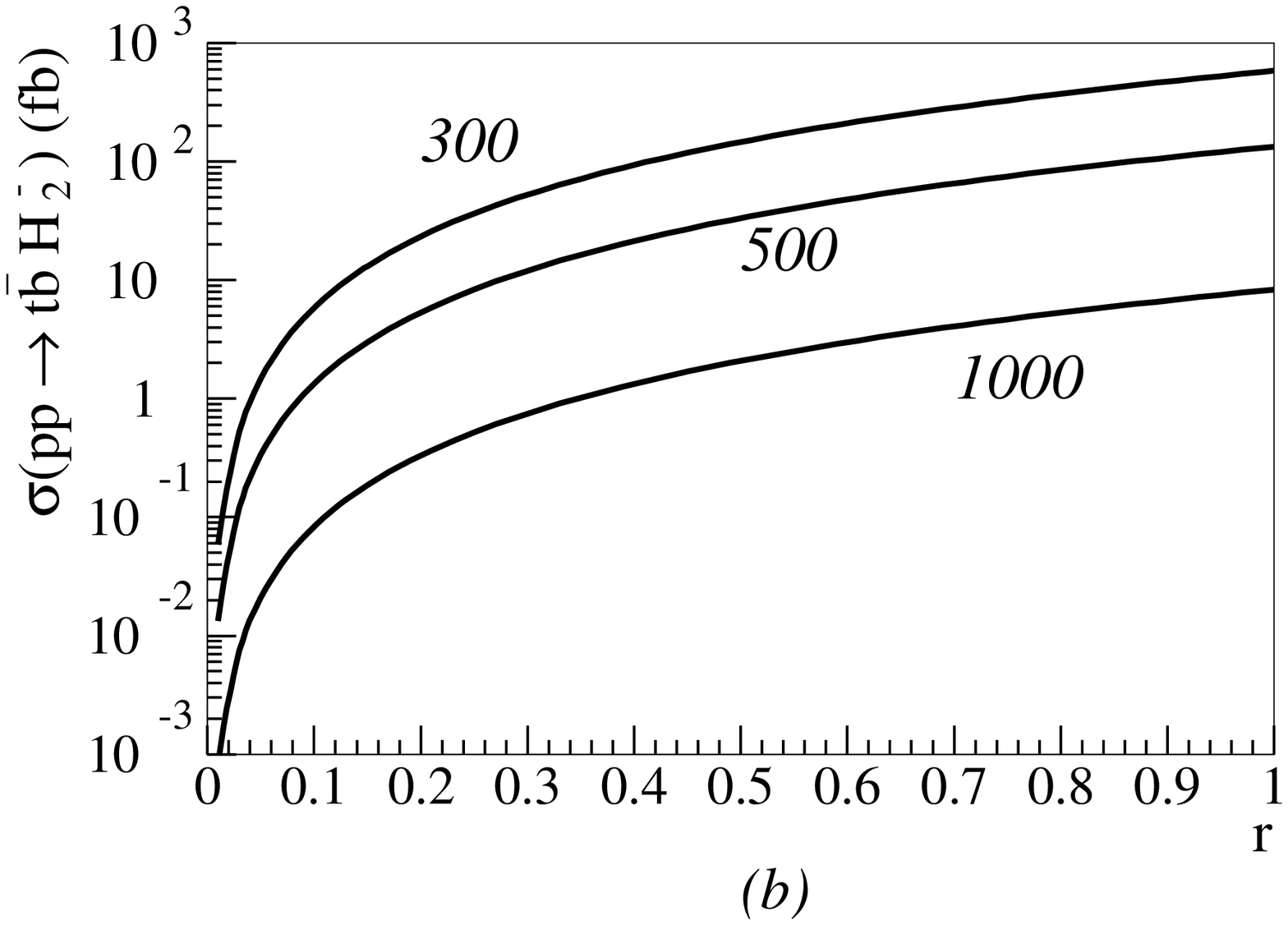}
\caption{Dependence of the cross section for associated $(a)$ $t\overline{b}H_{1}^{-}$ production with tan$\beta _T$ and $(b)$ $t\overline{b}H_{2}^{-}$ production with the ratio $r=B^D_L/C_{\beta _T}$ for $M_{H_{1,2}}=300, 500$ and $1000$ GeV}
\label{fig-3}
\end{figure}

Figs. \ref{fig-1}$(b)$-$(f)$ corresponds to single Higgs production in association with $\overline{b}t$ quarks (we show the production of the $H^{-}_{1,2}$ component. The $H^{+}_{1,2}$ is identical but in association with $b\overline{t}$ quarks). In this case, in addition to (\ref{H-H-V-couplings}), we must take into account the parameters from couplings in (\ref{L-neutro}) and (\ref{bottom-top-a}). Thus, we have the new free parameter $B^{D}_L$ for the quark couplings with $H_2^{\pm}$. This parameter corresponds to the mixing rotations associated with the down-type mass matrices $M_{DJ}$ from (\ref{mixing-mass}). According to Eq. (\ref{solution-L}), and for the third Down-type family, we obtain the following relation:

\begin{eqnarray}
B^D_{L}&\approx& \left(m_{b}\widetilde{S}^{\dagger}+\widetilde{s}m_{J}^{\dagger}\right)\left(m_{J}^2 \right)^{-1}%=m_b\left(\frac{\widetilde{S}^{\dagger}}{m_{J}^2}\right)+\frac{\widetilde{s}}{m_{J}},
\label{mixing_angle}
\end{eqnarray} 
where $m_b \approx 4$ GeV corresponds to the mass of the $b$ quark, $m_J \sim \upsilon _{\chi} \gg$ GeV is the mass of the heavy $J$ quarks, $\widetilde{S}=S\sim \upsilon _{\chi}$ the heavy mixing component of $M_{DJ}$ in (\ref{mixing-mass}), and $\widetilde{s}=s\sim \upsilon _{\rho, \eta} \sim$ GeV the corresponding light mixing components. Due to the above order of magnitudes of the VEVs, the mixing rotations in (\ref{mixing_angle}) exhibits small values ($B^D_{L} \ll 1$). However, the $H_2^{-}$ couplings in Eq. (\ref{bottom-top-a}) depend on the ratio $B^D_{L}/C_{\beta _T}$. Thus, although the mixing rotations takes small values, the couplings with the Higgs $H_2^{\pm}$ may be appreciable for small $C_{\beta _T}$. Then, we use the parameter $r=B^D_{L}/C_{\beta _T}$ instead of $B^D_{L}$ for our analysis. First, we obtain the cross section for associated production of $H_1^{\pm}$ Higgs bosons. Fig. \ref{fig-3}$(a)$ exhibits the dependence of the cross section with the parameter $T_{\beta _T}$ for the three $H_1^{\pm}$ masses, where large $T_{\beta _T}$ values lead to large production cross sections. Second, for the $H_2^{\pm}$-boson production, we calculate the dependence of the cross section with the parameter $r$, as shown in Fig. \ref{fig-3}$(b)$. Since the Yukawa couplings is proportional to $r$, the cross sections increase with $r$. 

%To obtain the cross section in type-II models, we see from Eqs. (\ref{bottom-top-a}) and (\ref{bottom-top-b}) that the couplings $g_{H_2bt}^I$ and $g_{H_2bt}^{II}$ are related by:

%\begin{eqnarray}
%\frac{g_{H_2bt}^I}{g_{H_2bt}^{II}}=T_{\beta _T}.
%\end{eqnarray}
%Thus, the dependence of the ratio between the cross section $\sigma^I(btH_2^{\pm})$ for type-I model and $\sigma ^{II} (btH_2^{\pm})$ for type-II model with $T_{\beta _T}$ is as shown in Fig. \ref{fig-4}(b). As it is to expect, both cross sections are identical for $T_{\beta _T}=1$.

\subsection{$H_{1,2}^{\pm}tb$ production}

\begin{figure}[b]
\centering \includegraphics[scale=0.7]{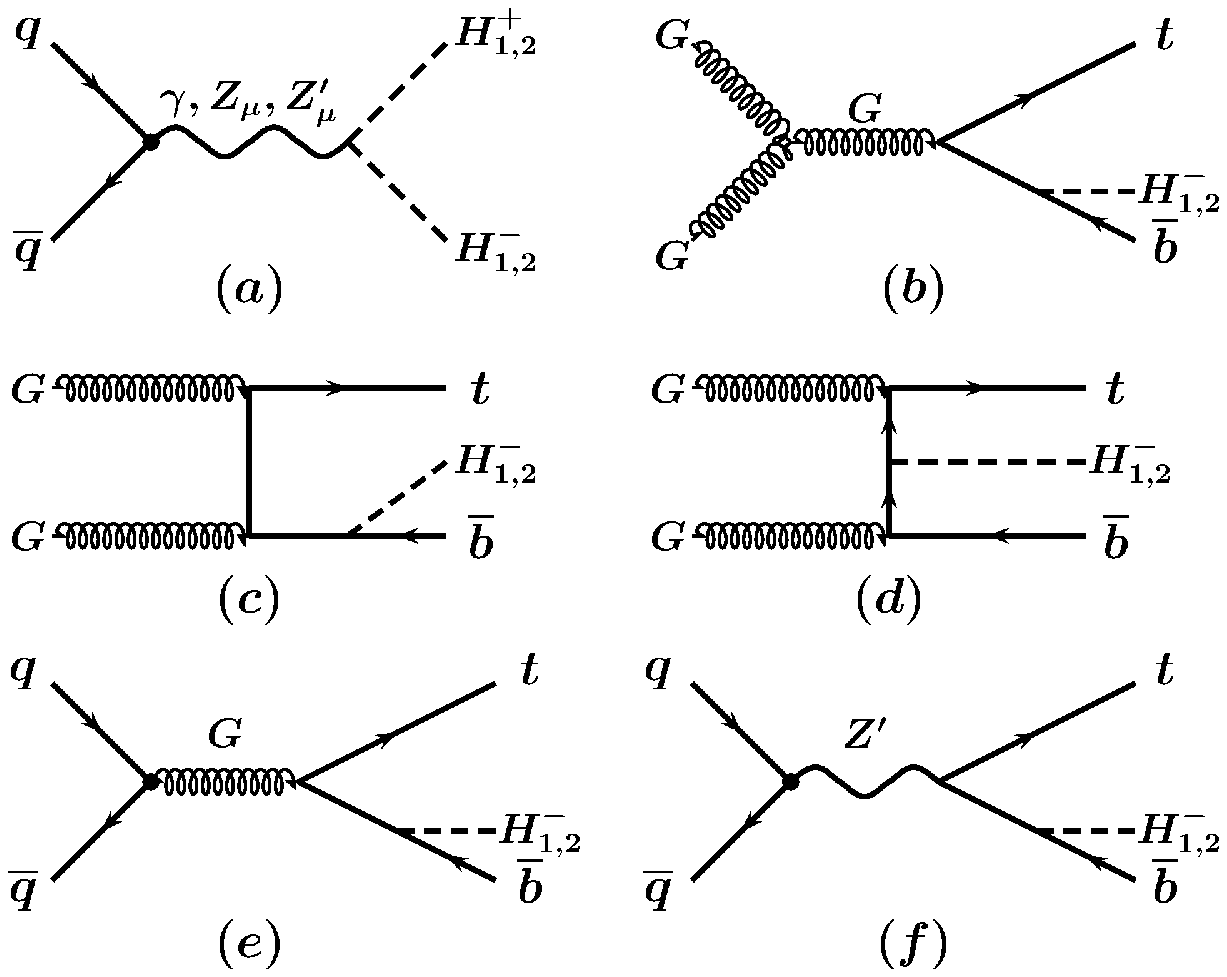}
\caption{Charged Higgs boson production in $pp$ collisions for $(a)$ pair production in quark-antiquark annihilation, $(b)$-$(d)$ associated single production in gluon-gluon collision and $(e)-(f)$ quark-antiquark annihilation}
\label{fig-1}
\end{figure}

\begin{figure}[b]
\centering \includegraphics[scale=0.45]{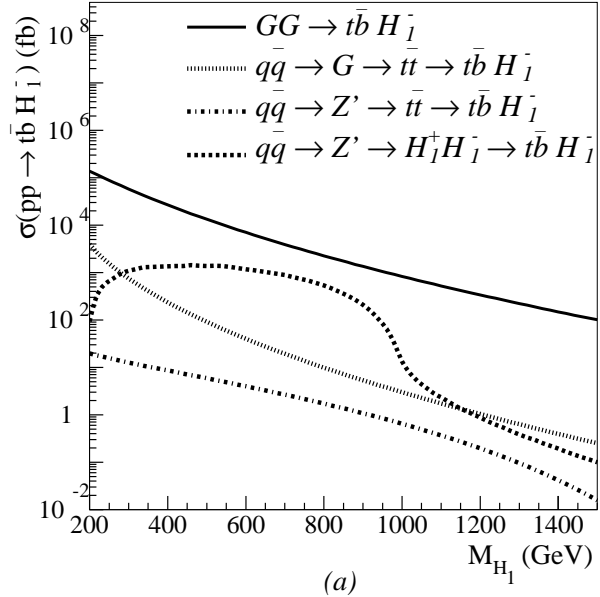}
\centering \includegraphics[scale=0.45]{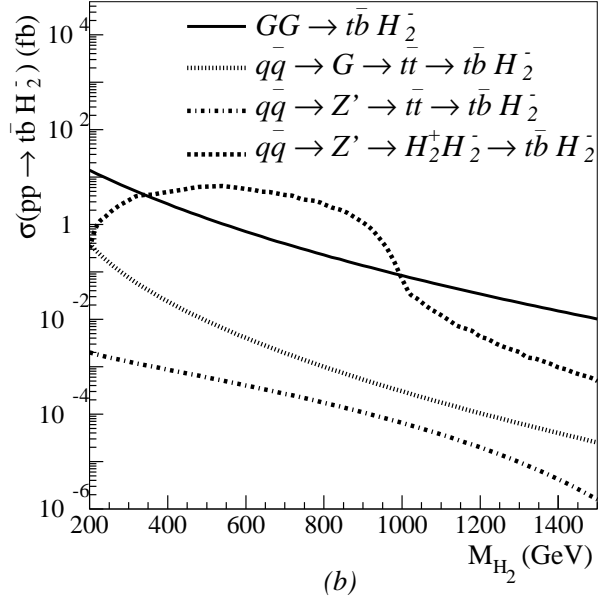}
\caption{Dependence of the cross section for $(a)$ $t\overline{b}H_{1}^{-}$ and $(b)$ $t\overline{b}H_{2}^{-}$ productions with the Higgs mass in different production modes. We use the inputs tan$\beta _T=10$, $r=0.1$ and $M_{Z'}=2$ TeV.}
\label{fig-4}
\end{figure}

In the above subsection we obtain the cross sections for single production in association with $bt$ quarks. However, we can obtain the same final states from pair production in the diagram \ref{fig-1}$(a)$ if we suppose that one of the Higgs bosons decay to $bt$. Thus, the total cross sections $\sigma (pp \rightarrow tbH_{1,2}^{\pm})$ have contributions from all the modes shown in Fig. \ref{fig-1}. To explore how large is the contribution from each mode, we calculate the corresponding cross sections using the following values: $M_{Z'}=2$ TeV, $T_{\beta _T}=10$ and $r=0.1$ (we note that these values are consistent with the condition of $B^{D}_L$ small: $B^D_L=rC_{\beta _T} \approx 0.01 \ll 1$). Figs. \ref{fig-4}$(a)$ and $(b)$ display the dependence of the cross sections with the Higgs masses for $H_1^{\pm}$ and $H_2^{\pm}$, respectively. We see that, in general, the dominant modes come from Gluos-Gluon collisions (Figs. \ref{fig-1}$(b)$-$(d)$). We also see that, as it is to expect, the contribution from Higgs pair production $q\overline{q} \rightarrow Z' \rightarrow H_{1,2}^{+}H_{1,2}^{-} \rightarrow t\overline{b}H_{1,2}^{-}$, becomes resonant for $M_{H_{1,2}}=M_{Z'}/2=500$ GeV. Furthermore, this mode becomes larger than the gluon collision for $H_2^{\pm}$ masses around the resonant region (see Fig. \ref{fig-5}$(b)$). The mode from Fig. \ref{fig-1}$(f)$ is, on the contrary, negligible for all the range of $M_{H_{1,2}}$.

\section{Decay of the charged Higgs bosons to leptons} 

To explore possibilities to distinguish different like-charged Higgs bosons, we study the decay mode $H^{\pm}_{1,2} \rightarrow \tau \nu$ from associated $H^{\pm}_{1,2}bt$ production. To have this decay, we first must study the lepton couplings in the Yukawa Lagrangian. From the spectrum in (\ref{fermion_spectrum}) and (\ref{331-scalar}), assuming Majorana mass terms for the singlets $N_R$, and according with the discrete symmetries in (\ref{discrete_sym}), we obtain the following renormalizable Yukawa Lagrangian for the lepton sector: 

\begin{eqnarray}
-\mathcal{L}_Y^{lep}&=&\overline{L^{i}_L}\left[\left(\chi h^N_{\chi ij}+ \eta h^N_{\eta ij}\right)N_R^j+ \rho h^e_{\rho ij}e_R^j\right]\notag \\
&+&\frac{1}{2}\sum_{\alpha, \beta, \gamma =1}^{3} \overline{L^{i(\alpha )}_L} \left(L^{j(\beta )}_L\right)^c h_{\rho} \rho^{*}_{(\gamma )}\epsilon ^{\alpha \beta \gamma }+\frac{1}{2}M_{Rij}\overline{N^{i}_R}N_R^{jc}+h.c,
\label{lepton_yukawa}
\end{eqnarray}
where $i,j=1,2,3$ label the family index (three $L_L$ triplets), $\alpha, \beta, \gamma =1,2,3$ label the flavor components into each triplet, and $\epsilon ^{\alpha \beta \gamma}$ corresponds to the antisymmetric Levi-Civita tensor ($1$ for even permutations of $(123)$, $-1$ for odd permutations, and $0$ if any index is repeated). For the charged sector, we obtain the following mass matrix after the symmetry breaking:

\begin{eqnarray}
M_e&=&\frac{1}{\sqrt{2}}h_{\rho}^e\upsilon _{\rho}.
\label{lepton_mass} 
\end{eqnarray}
For the neutral leptons, the Lagrangian in (\ref{lepton_yukawa}) contains mixing terms which produce neutrino mass terms. The authors in ref. \cite{catano} study three see-saw scenarios in order to obtain small neutrino masses. On the other hand, as with the quark Lagrangian, we rotate the Higgs sector to mass eigenstates according to Eqs. (\ref{331-mass-scalar-a}) and (\ref{331-mass-scalar-b}). In particular, for the couplings with the charged Higgs bosons and the SM leptons $(e^i,\nu ^i)$, we obtain:

\begin{eqnarray}
-\mathcal{L}_Y^{H_{1,2}^{\pm}}=\overline{\nu ^i_{L}}\left(-h_{\rho ij}^eS_{\beta _T}\right)e_R^{j}H_1^{+}+\overline{e^i_{L}}\left(-h_{\rho}\right)\nu ^{jc}_RH_2^{-}+h.c.
\label{charged-lept-yuk}
\end{eqnarray}

Using Eq. (\ref{lepton_mass}), we can write the coupling constant $h_{\rho}^e$ in terms of the lepton masses, while $\upsilon _{\rho} =\upsilon C_{\beta _{T}}$. Thus, for the tau lepton $(\tau )$, Eq.(\ref{charged-lept-yuk}) reads:

\begin{eqnarray}
-\mathcal{L}_Y^{H_{1,2}^{\pm}}=\frac{-\sqrt{2}m_{\tau}T_{\beta _{T}}}{\upsilon }\overline{\nu _{\tau L}}\tau _RH_1^{+}-h_{\rho}\overline{\tau _{L}}\nu ^{c}_{\tau R}H_2^{-}+h.c.
\label{tau-yuk}
\end{eqnarray}

In order to obtain small neutrino masses, the see-saw mechanism in 3-3-1 models requires small values of the parameter $h_{\rho}$ (typically $h_{\rho} \sim 10^{-7}-10^{-4}$). However, in the framework of the inverse see-saw mechanism, it is possible to obtain couplings as large as $h_{\rho }\sim 1$GeV$/\upsilon _{\rho}$ which is consistent with small neutrino masses \cite{catano}.
  
With the above considerations, we calculate the cross section distribution in $pp$ colisions in the CERN LHC hadron collider, based on an integrated luminosity $L=100 $ fb$^{-1}$ at CM energy of 14 TeV. We assume that $Br(H^{\pm}_{1,2}\rightarrow \tau \nu)=100$\%. We also consider the dominant contribution from gluon-gluon collisions (Figs. \ref{fig-1}$(b)$-$(d)$), and keep large $Z'$ masses to neglect its effects. We fix the following parameters: 

\begin{eqnarray}
T_{\beta _T}&=&10 \notag \\
r&=&0.1 \notag \\
h_{\rho }&\sim &1GeV/\upsilon _{\rho}=1GeV/\upsilon C_{\beta _{T}}\approx 0.041.
\end{eqnarray}

\begin{figure}[b]
\centering \includegraphics[scale=0.4]{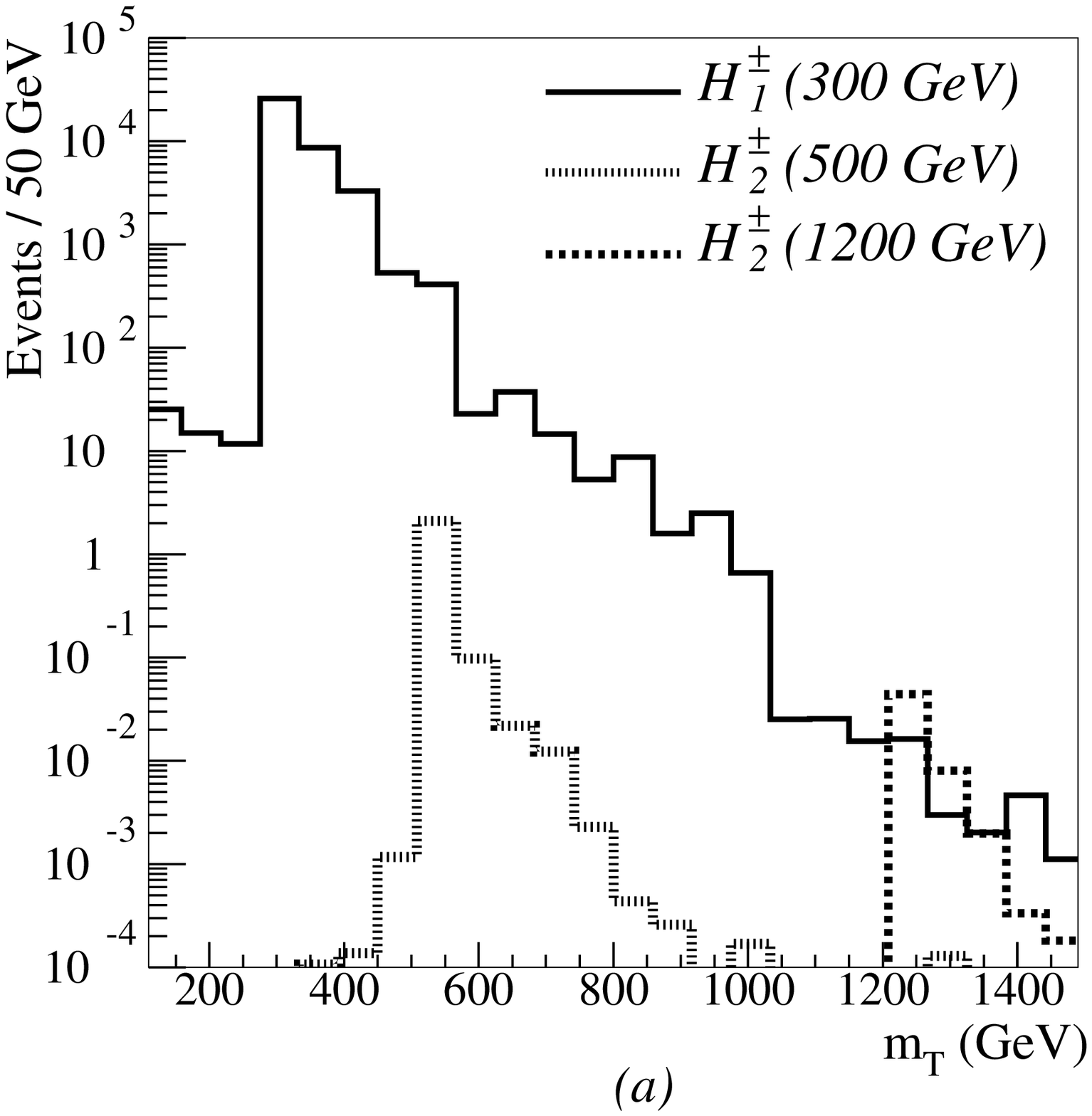}
\centering \includegraphics[scale=0.4]{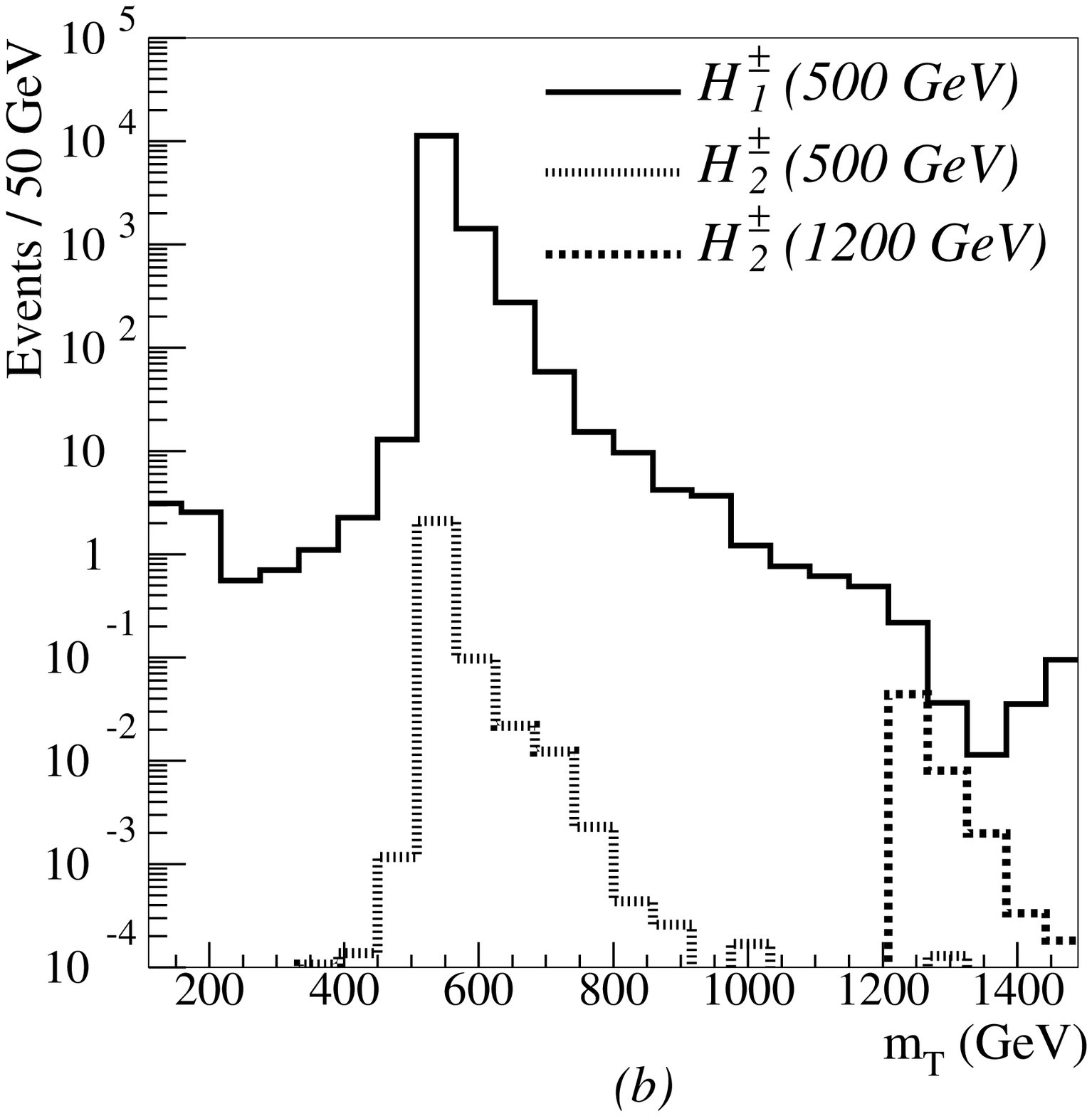}
\centering \includegraphics[scale=0.4]{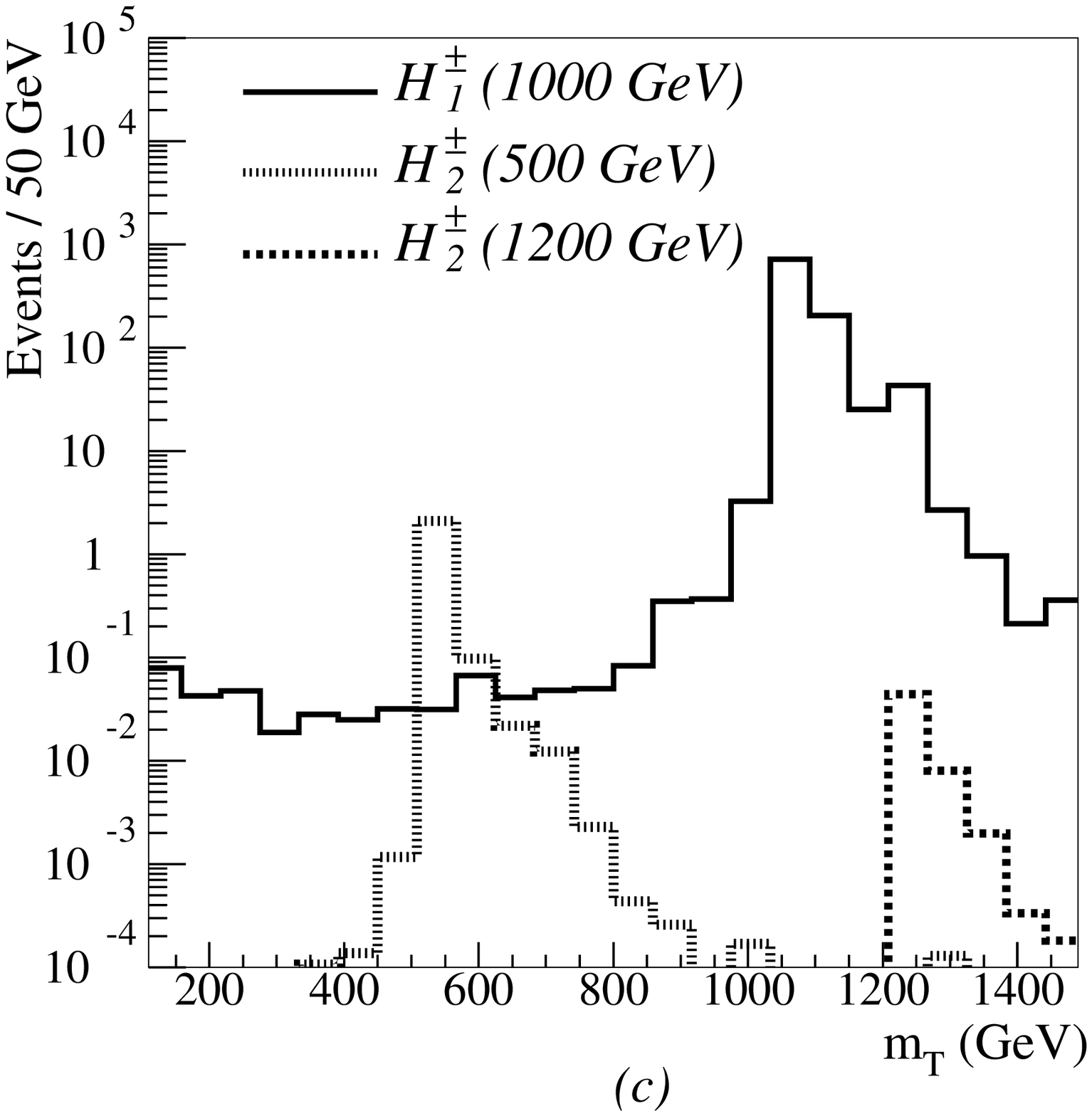}
\caption{Transverse mass distributions in $H_{1,2}^{\pm} \rightarrow \tau \nu _{\tau}$ decay in associated Higgs-boson production for $(a)$ $M_{H_1}=300$ GeV, $(b)$ $M_{H_1}=500$ GeV, and $(c)$ $M_{H_1}=1000$ GeV compared with $H_2^{\pm}$ bosons for $M_{H_2}=500$ and $1200$ GeV.}
\label{fig-5}
\end{figure}

Fig. \ref{fig-5} shows the Higgs boson tranverse mass distributions for the $\tau$ final state. The distribution in Fig. \ref{fig-5}$(a)$ shows the Higgs events for $M_{H_1} = 300$ GeV, and for $M_{H_2} = 500$ and $1200$ GeV. For $H_2^{\pm}$ bosons with mass 500 GeV, the signal is overlapped by the $H_1^{\pm}$ background, while for 1200 GeV bosons we see a small excess over the background. Fig. \ref{fig-5}$(b)$ shows the same distributions but for $M_{H_1} = 500$ GeV, which overlaps the $H_2^{\pm}$ signals. Finally, Fig. \ref{fig-5}$(c)$ displays distributions for $M_{H_1} = 1000$ GeV, where an observable $H_2^{\pm}$ signal with mass 500 GeV arises over the $H_1^{\pm}$ background. Although the production of the hipercharge-two Higgs boson is in general small in relation with the hypercharge-one Higgs, we see from the above analysis possible scenarios where the two signals may be distinguishable. Furthermore, this small signal could be improved if more sophisticated discriminating distributions are used, or other decay channels are considered \cite{gross}. 

\section{Conclusions}          
 
The identification of multiple Higgs boson signals could reveal many features about the underlying model beyond the SM. For example, if a charged Higgs boson is detected, further analysis will be necessary to test the compatibility of different models with the experimental data. In this paper we show that both versions of the 2HDM (type-I and type-II models), which contains one charged Higgs boson, may be embedded into a 3-3-1 model, which exhibits two charged Higgs bosons: a hipercharge-one $H_1^{\pm}$ and a hypercharge-two $H_2^{\pm}$ Higgs boson. At low energy, $H_1^{\pm}$ can be identified with the charged Higgs boson of the 2HDM, while $H_2^{\pm}$ are other bosons from the underlying 3-3-1 model. Thus, in this case, the identification of two like-charged Higgs boson signals may reveal new physics beyond 2HDM. Taking into account mixing couplings between the two scales of the model ($\upsilon \sim $GeV and $\upsilon _{\chi } \gg $GeV), we show that the $H_2^{\pm}$ can be produced through the same production channels as the $H_1^{\pm}$. Using the method of recursive expansion, we found that after rotations to mass eigenstates, Yukawa couplings between the SM fermions and $H_2^{\pm}$ arise due to the small mixing angle associated to the rotation matrix $B_L^{D}$. Furthermore, since the Yukawa couplings appear through the ratio $B^{D}_L/C_{\beta _T}$ for type-I and -II models, events of $H_2^{\pm}$ may be enhanced to observable scales if $C_{\beta _T}$ takes small values. We show that pair production of charged Higgs bosons can be significantly enhanced due to the contribution of a heavy $Z'$ neutral gauge boson, predicted by the model. The dominant mode for associated $btH^{\pm}_{1,2}$ is through gluon-gluon collisions. However, pair production through quark anhilation can be as large as single production due to resonant intercharge of the heavy $Z'$ boson in 3-3-1 models. By considering decays to leptons $H_{1,2}^{\pm} \rightarrow \tau \nu _{\tau}$, we obtain scenarios where small peaks of $H_{2}^{\pm}$-boson events in transverse mass distributions can be identified over the $H_{1}^{\pm}$ background. This small signal could be improved if more sophisticated discriminating distributions are used. Thus, in case that charged Higgs bosons are detected, the identification of multiple like-charged Higgs boson signals may be a possible discriminating method to test different theoretical models beyond the SM.

This work was supported by Colciencias.

%\begin{figure}[b]
%\centering \includegraphics[scale=0.45]{fig_4a.eps}
%\centering \includegraphics[scale=0.45]{fig_4b.eps}
%\caption{$(a)$ Dependence of the cross section for associated $t\overline{b}H_{2}^{-}$ production with the ratio $r=B^D_L/C_{\beta _T}$ for $M_{H_{2}}=300, 500$ and $1000$ GeV in type-I models. $(b)$ Dependence of the cross section ratio between type-I and -II models with tan$\beta _T$.}
%\label{fig-4}
%\end{figure}

\end{document}